\documentstyle[12pt,amssymb,latexsym,amsmath]{article}
\newcommand{\NN}{{\mathbb N}}

\newcommand{\RR}{{\mathbb R}}

\newcommand{\beq}{\begin{equation}}
\newcommand{\eeq}{\end{equation}}
\newcommand{\ba}{\begin{array}}
\newcommand{\ea}{\end{array}}
\begin{document}

\title{Dressing method based on homogeneous Fredholm equation: 
 quasilinear PDEs in multidimensions}

\author{
A.I. Zenchuk\\
Center of Nonlinear Studies of L.D.Landau Institute
for Theoretical Physics  \\
(International Institute of Nonlinear Science)\\
Kosygina 2, Moscow, Russia 119334\\
E-mail: zenchuk@itp.ac.ru
\\\\
P.M. Santini\\Dipartimento di Fisica, Universit\`a di Roma "La Sapienza" \\
and Instituto Nazionale di Fisica Nucleare, Sezione di Roma1, \\
Piazz.le Aldo Moro 2, I-00185 Roma, Italy\\
E-mail: paolo.santini@roma1.infn.it
}

\maketitle
%%%%%%%%%%%%%%%
\begin{abstract}
In this paper we develop a dressing method for constructing and solving some 
classes of matrix quasi-linear Partial Differential Equations (PDEs) in arbitrary dimensions. 
This method is 
based on a homogeneous integral equation with a nontrivial kernel, which allows one to 
reduce the nonlinear PDEs to systems of non-differential (algebraic or transcendental) 
equations for the unknown fields.  In the 
simplest examples, the above dressing scheme captures matrix equations integrated by the 
characteristics method and nonlinear PDEs associated with matrix  Hopf-Cole transformations. 
\end{abstract}

%%%%%%%%%%%%%%%

\section{Introduction}
Since the pioneering work \cite{GGKM} on the Korteweg - de Vries equation \cite{KdV}, 
completely integrable nonlinear Partial Differential 
Equations (PDEs) have been intensively studied during the 
last decades, and large classes of integrable PDEs have been found, 
like the so called $S$-integrable systems (or soliton equations, whose integration scheme  
involves the solution of a linear integral equation) \cite{ZMNP,AS},  
 and the so-called $C$-integrable equations (integrated by 
simpler transformations, like the Hopf-Cole transformation for the Burgers equation) \cite{Calogero}. 
Much effort has been devoted to the study 
of direct techniques to construct and solve nonlinear PDEs. One of the most 
powerful of such techniques is the dressing method, 
originally developed for (1+1) and (2+1)-dimensional $S$-integrable  
models \cite{ZSh1,ZSh2,ZM,BM} (see also \cite{Konop}). Multidimensional  
generalizations of it have also been developed \cite{Zakharov1,Zakharov2,Zakharov,Z},  
allowing to integrate  special classes of higher dimensional nonlinear PDEs.
Nonlinear dressing methods for PDEs in arbitrary dimensions associated with vector fields  
are also known \cite{BK,MS}. All the above dressing formalisms are based on the hypothesis that,
for given spectral data, the spectral function can be uniquely constructed from the 
relevant integral equation; i.e., the kernel of the corresponding integral 
operator is empty. 

Recently, a new version of the dressing method has appeared, based on 
integral operators with nontrivial kernel \cite{ZS}. This 
assumption removes most of the restrictions on the dimensionality of the  
space of analytic solutions of the constructed  PDEs, and the solutions of the  
 $n$-dimensional PDEs  constructed in \cite{ZS} are parametrized by arbitrary spectral 
functions of $(n-2)$ variables (full integrability would be achieved if the variables 
were $(n-1)$). The structure of the nonlinear PDEs has been 
simplified and the space of analytic solutions has been enriched in \cite{Z3}, where  
multidimensional dressing operators \cite{Z2}, which are not effective in the classical case, 
have been used. The solutions of the $n$-dimensional nonlinear PDEs constructed there 
are parametrized by arbitrary spectral functions of $(n-1)$ variables, but full integrability 
has not been achieved even there, since these spectral functions are not in the right number. 
 
In this paper we develop another variant of the dressing method, based on a homogeneous 
integral equation with nontrivial kernel, allowing one to reduce certain classes of nonlinear 
PDEs in arbitrary dimensions to systems of non-differential (algebraic and/or transcendental) 
equations, similarly to the method of characteristics \cite{Whitham}. The nonlinear PDEs in 
arbitrary dimensions isolated by this method are built in terms of ``dressed'' first order 
operators of the type (\ref{L_m}) below. To increase the dimensionality of these  
PDEs and of their space of analytic solutions, we use  
multidimensional differential operators first introduced in \cite{Z3}. 

Below is the list of the nonlinear PDEs which will be derived and solved in this paper. 
All equations 
are $Q\times  Q$ matrix equations ($Q\in\NN_+$), unless differently specified.  
Hereafter we write superscripts inside of parenthesis in order to distinguish them from the power 
notation.

1. The first order matrix equation
\begin{eqnarray}\label{char}
 w_{t_1} + \sum_{j=1}^N  w_{x_j} \rho^{(j)}(w) = [w,T \rho^{(0)}(w)] 
\end{eqnarray}
will be derived in  Sec.\ref{First_order}. Here $T$ is any constant matrix, $\rho^{(i)}(w)$ are 
arbitrary matrix functions representable as positive 
power series of $w$, and $N$ is any integer. This matrix equation can also be integrated 
\cite{SZ} by the method of characteristics. 

2. A minor modification of the dressing method for eqs.(\ref{char}) allows one 
to construct a class of second 
order matrix systems (see Sec.\ref{Second order PDEs}) which degenerate, for the simplest 
choice of the arbitrary functions, to the matrix Burgers equation, linearizable by the Hopf-Cole  
transformation. A typical example 
is given by the second order matrix system:
\begin{eqnarray}\label{2_nl1}\label{ex2}
\ba{l}
{\cal{L}}_2(w)=[w,S(v^2-{\cal{L}}_1(v))v]
, \\
{\cal{L}}_2(v) +  S{\cal{L}}_1^2 (v)  = 
S{\cal{L}}_1(v^2) + [ S {\cal{L}}_1 (v)  , v] + [ v, Sv^2], 
\ea
\end{eqnarray}
subjected to the constraint
\beq\label{A00}
\ba{l}
{\cal{L}}_1(w)=[w,v], 
\ea
\eeq
where $S$ is any constant diagonal matrix, the    
differential operators ${\cal{L}}_m,\;m=1,2,$ are defined as follows:
\begin{eqnarray}\label{L_m}
{\cal{L}}_m(f(x)) =f_{t_m}(x) + 
\sum_{j=1}^N f_{x_j}(x) \rho^{(mj)}(w),
\end{eqnarray}
and $\rho^{(mj)}(w)$ are 
arbitrary matrix functions representable as positive  
power series of $w$.

We remark that many
equations of Mathematical Physics appear as first order
quasilinear PDEs in multidimensions. Therefore it is important to 
develop efficient methods to isolate and solve, 
among these equations, integrable and/or partially integrable cases.   

The matrix equation (\ref{char}), a remarkable example of integrable system 
of PDEs in arbitrary dimensions, is the natural matrix generalization of the 
classical examples of physically relevant scalar equations integrable by the method of
characteristics. A systematic study of the matrix reductions of (\ref{char}), 
with the goal of isolating physically relevant cases, will be the subject 
of a subsequent paper. Here we only remark that, in the $2\times 2$ matrix reduction 
\beq
w=\left(
\ba{cc}
\nu_1 & \nu_2 \\
\nu_2 & \nu_1
\ea
\right),
\eeq
equation (\ref{char}), with $N=1$ and $\rho^{(0)}=0$, coincides \cite{SZ} with the gas-dynamics 
equations \cite{Whitham}
\beq \label{gas}
\ba{l}
{\nu_1}_t+\nu_1{\nu_1}_{x_1}+\nu_2{\nu_2}_{x_1}=0, \\
{\nu_2}_t+\nu_2{\nu_1}_{x_1}+\nu_1{\nu_2}_{x_1}=0.
\ea
\eeq
We also remark that other vector generalizations of scalar equations  
integrable by the method of
characteristics have been studied during the last twenty years, by different methods, 
in a set of papers \cite{ts1,dn,ts2}.

The system of second order equations (\ref{2_nl1}), constructed in terms of the two first order 
multidimensional operators (\ref{L_m}), presents mathematical features 
in common with equations integrable by the method of characteristics, and with equations integrable 
by the Hopf-Cole transformation; therefore it is   
conceptually similar to the equations, introduced in \cite{Z}, presenting mathematical features 
of both $S$ - integrable and $C$ - integrable PDEs. Equations like (\ref{2_nl1}), for which integrability 
properties of different type merge together, are interesting prototype examples    
in the theory of integrable systems. 

The paper is organized as follows. In Sec. \ref{General}, after a brief review of the classical 
dressing method (in Sec.2.1) and of the novel dressing features contained in \cite{ZS} 
(in Sect.2.2), we 
describe the main features of the general dressing algorithm of this paper.  
Eqs.(\ref{char})  will be derived  in Sec.\ref{First_order}.
In Sec.\ref{Solutions} we use the dressing algorithm to reduce Eqs.(\ref{char}) to a 
system of non-differential equations characterizing the general solution of (\ref{char});  
in particular, we discuss the Cauchy problem.  In Sec.\ref{Second order PDEs}
we derive eqs.(\ref{ex2}) and the associated general class, discussing their solution space.
In Sec.\ref{Linear} we make some remarks on the overdetermined system of linear PDEs for the 
associated spectral function.  Conclusions are presented in Sec.\ref{Conclusions}. 

%%%%%%%%%%%%%%%%%%%%
%%%%%%%%%%%%%%
\section{Basic novelties of the dressing method}
\label{General}

To emphasise all the significant  novelties
and features of our algorithm, in comparison with both the classical dressing method 
and the new dressing 
method developed in \cite{ZS},  first, we give a brief review of one of the versions of the 
classical dressing method for the (2+1)-dimensional $N$-wave equation (Sec.\ref{Review_class}), 
and, second, we describe the modifications introduced in \cite{ZS} (Sec.\ref{Review_17}). 
After such a preliminary overview, we explain the novelties of the algorithm introduced in
this paper, (Sec.\ref{Novelties}).

%%%%%%%%%%%%%%%%%%%%%%%%
\subsection{Brief review of the classical version of the dressing method, \\ 
and the (2+1)-dimensional $N$-wave equation}
\label{Review_class}
The starting point of one of the versions of the classical dressing method is the linear integral 
equation
\begin{eqnarray} \label{Sec1:U}
\Phi(\lambda;x)=\int \Psi(\lambda,\mu;x)U(\mu;x)d\Omega(\mu)
\equiv \hat\Psi U(\lambda;x),
\end{eqnarray}
in the spectral variables $\lambda=(\lambda_1,\dots,\lambda_{\dim\lambda}),~\mu=(\mu_1,\dots,\mu_{\dim\lambda})$, 
for the unknown matrix function $U$. The given matrix functions $\Phi$ and $\Psi$ are defined by 
some extra conditions, which fix their dependence on an additional vector parameter $x= (x_1,\dots,x_{\dim x})$, 
whose components are the independent variables of the associated nonlinear PDEs. $\Omega$ is some largely arbitrary 
scalar measure in the $\mu$-space. Apart from $\Omega$, all the functions appearing in this paper 
are $Q\times Q$ matrix functions.

We remark that no a priory assumption is made in (\ref{Sec1:U}) 
on the dependence of $\Psi$ on $\lambda$ 
(this general starting point has been used, for instance, in 
\cite{SAF} and in \cite{Z}), to keep the structure 
of $\Psi$ as much general as possible. Indeed, although in 
most of the cases such a dependence is described by a Cauchy kernel, an  
indication that equation (\ref{Sec1:U}) is the manifestation of 
Riemann-Hilbert and/or $\bar\partial$ analyticity problems, 
there are examples (see \cite{SAF} and \cite{Z}) in which       
more general representations appear, 
indicating that the above analyticity 
problems could be a too restrictive starting points.

%%%%%%%%%%%%%%%%%%%%%%%%%%%%%%%%%%%%%%%%%%%
\subsubsection{Derivation of $N$-wave system.}
\label{Section:class}

The basic assumption underlying all the known classical 
dressing procedures available in the literature, is that the operator $\hat\Psi$ 
in (\ref{Sec1:U}) is uniquely invertible; i.e., that
\begin{eqnarray}\label{Sec1:nondeg}
\dim {\mbox{ker}} \hat\Psi =0.
\end{eqnarray}
The $x$-dependence is introduced by the matrix equations 
\begin{eqnarray} \label{Sec1:x}
\Psi_{x_i}(\lambda,\mu;x)=\Phi(\lambda;x)B_iC(\mu;x),~~~i=1,..,{\mbox{dim}}~x,
\end{eqnarray}
showing that the $x$-derivatives of the kernel $\Psi$ are degenerate matrix functions, another basic feature 
of all known classical dressing algorithms,   
where $B_i,~i=1,..,\dim~x,$ are  constant diagonal matrices, at
most $Q$ of them may be independent. Due to the above degeneracy, the compatibility 
of equations (\ref{Sec1:x}) leads to separate equations for $\Phi$ and $C$: 
\begin{eqnarray} \label{Sec1:Phi_x}
\Phi_{x_i} B_j - \Phi_{x_j} B_i = 0,~~~~i\ne j,\\\label{Sec1:c_x}
B_j C_{x_i} - B_i C_{x_j} = 0,~~~~i\ne j,
\end{eqnarray}
and one equation is the adjoint of the other. Without loss of generality we assume 
$B_1=I$, where $I$ is the identity matrix.

Replacing, in equation (\ref{Sec1:Phi_x}), $\Phi$ by $\hat\Psi U$, as indicated in (\ref{Sec1:U}), and using 
(\ref{Sec1:x}), one obtains the following equation:
\begin{eqnarray}
\label{Sec1:PsiU}
\hat\Psi L_{ij}U=0, 
\end{eqnarray}
where
\begin{eqnarray}
\label{Sec1:LUU}
L_{ij}U\equiv U_{x_i} B_j - U_{x_j} B_i+ UB_i v B_j - U B_j v B_i,~~i,j=1,..,{\mbox{dim}}~x,~~~i\ne j
\end{eqnarray}
and 
\begin{eqnarray}
\label{Sec1:v}
v(x)\equiv \int C(\lambda;x)U(\lambda;x)d\Omega(\lambda).
\end{eqnarray}
Then the property (\ref{Sec1:nondeg}) implies that 
 \begin{eqnarray}\label{Sec0:UU}
 L_{ij} U(\lambda;x)=0,~~i,j=1,..,{\mbox{dim}}~x,~~~i\ne j
 \end{eqnarray}
 or, explicitly:
\begin{eqnarray}\label{Sec0:U_lin}
L_{21}U=U_{x_2}   - U_{x_1} B_2 - U [v, B_2]=0,\\\nonumber
L_{31}U=U_{x_3}   - U_{x_1} B_3 - U [v, B_3]=0,\\\nonumber
\end{eqnarray}
having chosen $j=1$, $i=2, 3$. 

This is nothing but the well-known linear overdetermined 
system corresponding to the $N$-wave equation in the three variables $x_1,x_2,x_3$. 

\iffalse
The associated complete system of nonlinear PDEs is simply obtained, in the dressing philosophy, 
``saturating the parameter $\lambda$'' in equations (\ref{Sec0:U_lin}) by the integral operator 
$\int d\Omega(\lambda) C(\lambda;x)\cdot$:
\begin{eqnarray}\label{Sec0:nl}
L_{21} v-[B_2,v^1]=v_{x^2} - v_{x^1} B_2 - v [v, B_2] 
-[B_2, v^1]  =0,\\\nonumber
L_{31} v-[B_3,v^1]=v_{x^3} - v_{x^1} B_3 - v [v, B_3] 
-[B_3, v^1]  =0.
\end{eqnarray}
It is written in terms of the square matrix fields $v(x)$ and $v^1(x)$, where
\begin{eqnarray}\label{Sec0:v1}
v^1(x)\equiv \int C_{x^1}(\lambda;x)U(\lambda;x)d\Omega(\lambda).
\end{eqnarray}
Eliminating $v^1$ from these two equations, we get the celebrated $N$-wave system in $3$ dimensions:
\fi
Its compatibility condition yields
\begin{eqnarray}\label{Sec0:Nw1}
[v_{x_3},B_2] - [v_{x_2},B_3] + B_2 v_{x_1} B_3 -  B_3 v_{x_1} B_2
-[[v,B_2],[v,B_3]]=0.
\end{eqnarray}

\iffalse
The same equation may be derived directly from the compatibility condition of the system
(\ref{Sec0:U_lin}).

Similarly, considering 
the equations $L_{j1}U=0$ and $L_{k1}U=0$ for any $ j \neq k \neq 1$, one  
derives the hierarchy of $n$-wave equations
 \begin{eqnarray}\label{Sec0:Nw2}
[v_{x^k},B_j] - [v_{x^j},B_k] + B_j v_{x^1} B_k -  B_k v_{x^1} B_j
-[[v,B_j],[v,B_k]]=0.
 \end{eqnarray}

We remark that, in the above dressing construction, the linear integral operator $\hat\Psi$ in (\ref{Sec1:U}) 
acts from left and, consequently, the partial differential operators $L_{ij}$ in (\ref{Sec0:UU}) 
act from right, while, in the soliton literature, one usually makes the opposite choice. 
\fi

It is important to remark that 
\begin{enumerate}
\item
eq.(\ref{Sec0:Nw1}) may be derived in a different way, ``saturating the parameter $\lambda$'' 
in equations (\ref{Sec0:U_lin}) by the integral operator 
$\int d\Omega(\lambda) C(\lambda;x)\cdot$, and obtaining the nonlinear system:
\begin{eqnarray}\label{Sec0:nl}
L_{21} v-[B_2,v_1]=v_{x_2} - v_{x_1} B_2 - v [v, B_2] 
-[B_2, v_1]  =0,\\\nonumber
L_{31} v-[B_3,v_1]=v_{x_3} - v_{x_1} B_3 - v [v, B_3] 
-[B_3, v_1]  =0,
\end{eqnarray}
written in terms of the square matrix fields $v(x)$ and $v_1(x)$, where
\begin{eqnarray}\label{Sec0:v1}
v_1(x)\equiv \int C_{x_1}(\lambda;x)U(\lambda;x)d\Omega(\lambda).
\end{eqnarray}
Eliminating the extra field $v_1$ from these two equations, we get the $(2+1)$-dimensional 
$N$-wave system (\ref{Sec0:Nw1}).

The possibility to derive integrable systems in these two alternative ways is important, since,  
while integrable PDEs in 2+1 dimensions (or less) are characterized as the compatibility condition of 
a linear overdetermined system of PDEs, such a basic property seems to be lost in multidimensions.
\item
Each linear equation (\ref{Sec0:U_lin}) is two-dimensional.
\end{enumerate}

%%%%%%%%%%%%%%%%%%%%%%%%%%%
\subsubsection{Solution space}
 
We now consider the manifold of analytic solutions of 
equation (\ref{Sec0:Nw1}) generated by the dressing procedure. 
The solutions of eqs.(\ref{Sec1:Phi_x}) and (\ref{Sec1:c_x}) can be parametrized as follows:
\begin{eqnarray} \label{Sec0:Phi}
\Phi(\lambda;x)= \int \Phi_0(\lambda,k)e^{kB\cdot x} d k,\\
\label{Sec0:c}
C(\mu;x)=\int e^{qB\cdot x} C_0(\mu,q) d q,
\end{eqnarray}
where
\begin{equation}
B\cdot x=\sum_{i=1}^{\dim x}B_i x_i,
\end{equation}
and where the spectral parameters $\lambda,\mu, k,q$ are scalars. 
It is simple to see, from the linear limit,  that the solution space of equation (\ref{Sec0:Nw1}), generated 
by the dressing algorithm, is full. Indeed, in the linear limit: 
$\Psi(\lambda,\mu) \sim \delta(\lambda-\mu)$ and $U\sim \Phi$. Take 
$C_0(\lambda,q) = \delta (\lambda- q)$;  
then the solution $v$ of the 3 - dimensional $N$-wave system (\ref{Sec0:Nw1}), which in the linear limit reads 
\begin{eqnarray}
\label{dim2}
v(x)\sim \int C(\lambda;x)\Phi(\lambda;x) d\Omega(\lambda) =
 \int e^{\lambda B\cdot x} \Phi_0(\lambda,k)e^{kB\cdot x} d kd\Omega(\lambda),
\end{eqnarray}
is parametrized by the arbitrary matrix function $\Phi_0(\lambda,k)$ of the two scalar spectral 
parameters $\lambda,k$; then its solution space is 2 dimensional, and therefore it is complete. 

\iffalse
We end this section remarking that the Cauchy kernel appearing 
in (\ref{Sec2:Psi}), obtained here as a consequence of 
equations (\ref{Sec0:Phi}),(\ref{Sec0:c}) and (\ref{Sec1:x}), 
is a manifestation of the distinguished analyticity 
properties of the solution $U(\lambda;x)$ in the complex $\lambda$ plane, 
in agreement with the well-known derivations of the $N$-wave equation  (\ref{Sec0:Nw1}) 
from Riemann-Hilbert and /or $\bar\partial$ problems \cite{Kaup,F,FA}. 
\fi

%%%%%%%%%%%%%%%%%%%%%%%%%%%%%%%%
\subsection{Novelties of the dressing methods  introduced in \cite{ZS}}
\label{Review_17}
In \cite{ZS} we assumed that the kernel of the operator $\hat\Psi$ is one dimensional:
\begin{eqnarray}\label{Sec2:deg}
\dim\mbox{ker}\hat\Psi =1,
\end{eqnarray}
i.e., the solution of the homogeneous equation associated with eq.(\ref{Sec1:U}) is nontrivial:
\begin{eqnarray}\label{Sec1:Unhom}
 0=\hat\Psi H \;\;\Leftrightarrow \;\; H(\lambda;x) = U^h(\lambda;x)f(x),
\end{eqnarray}
where $U^h(\lambda;x)$ is some nontrivial solution of the homogeneous equation $\hat\Psi H=0$ and 
$f(x)$ is an arbitrary matrix function of $x$. Then the general solution of eq.(\ref{Sec1:U}) reads
\begin{eqnarray}\label{Sec1:Uninhom}
U(\lambda;x)=U^p(\lambda;x) + U^h(\lambda;x) f(x),
\end{eqnarray}
where $U^p(\lambda;x)$ is some particular solution of (\ref{Sec1:U}).

As a consequence of the novel assumption (\ref{Sec2:deg}), equation (\ref{Sec1:PsiU}) implies the following equations
 for $U$:
\begin{eqnarray}\label{Sec1:LU}
{\cal{E}}_j(\lambda;x)\equiv  L_{j1} U(\lambda;x) - (L_{21} U(\lambda;x)) A^{j}(x)=0,\;\;\;j=3,\dots, \dim x, \\
L_{j1}U\equiv U_{x_j} -U_{x_1} B_j -U [v,B_j],\;\;\;\;j=2,\dots,\mbox{dim }x,
\end{eqnarray}
where $A^j(x)$ are some matrix functions to be defined. 
We have established that, if $\dim\mbox{ker}\hat\Psi =1$, then each 
linear equation (\ref{Sec1:LU}) for the spectral function $U(\lambda;x)$ is 3 dimensional. 

 The associated nonlinear equations are obtained ``saturating the parameter 
$\lambda$'' in equations (\ref{Sec1:LU}) by 
the integral operator $\int d\Omega(\lambda)  C(\lambda;x)\cdot$.
In order to express $A^{j}(x)$ in terms of $U$ and close the system, we introduce 
an external dressing function $G(\lambda;x)$, and the associated new matrix fields 
\begin{eqnarray}\label{Sec0:w^ij}
w^{00}(x)\equiv \int G(\lambda;x)U(\lambda;x)d\Omega(\lambda),\;\;
w^{j0}(x)\equiv \int G_{x^j}(\lambda;x)U(\lambda;x)d\Omega(\lambda),\;\;j>0, \\
w^{ij}(x)\equiv \int G_{x^ix^j}(\lambda;x)U(\lambda;x)d\Omega(\lambda),\;\;i,j>0,\;\;
w^{ij}(x)=w^{ji}(x).
\end{eqnarray}

Since the dimensionality of $G$ has no formal restrictions, the above $w$ - fields increase the dimensionality of the nonlinear PDEs. 
This can be seen, for instance, from their small field limits: $w^{00}(x)\sim \int G(\lambda;x) \Phi(\lambda;x) d\Omega(\lambda)$,   
$w^{j0}(x)\sim \int G_{x_j}(\lambda;x) \Phi(\lambda;x) d\Omega(\lambda)$.   
Nonlinear equations for these fields appear after applying the  
 integral operators $\int d\Omega(\lambda)  G(\lambda;x)\cdot$, $\int d\Omega(\lambda)  G_{x_j}(\lambda;x)\cdot$ to eq.(\ref{Sec1:LU}).
 
To close the system of nonlinear PDEs, one needs (a) equations defining $G(\lambda;x)$ and 
(b) an additional relation between all the matrix fields, which fixes the arbitrary function $f(x)$ in the 
solution space, see eq.(\ref{Sec1:Uninhom}), and may be taken in quite arbitrary form
\begin{eqnarray}\label{Sec1:condition00}
F(v,v^1,w^{00},w^{i0},w^{ij})=0,\;\;\;\;i,j=1,2,\dots.
\end{eqnarray}

Let us collect the basic novelties of the algorithm.
\begin{enumerate}
\item The existence of a nontrivial kernel of the 
basic integral equation implies that the solutions constructed by the dressing depend on an arbitrary function $f(x)$ of the 
coordinates; this fact has the following  important implications.
\item The nonlinear system of PDEs constructed by the dressing scheme possesses a distinguished 
block structure and is underdetermined. 
\item
To close the system and fix its underdeterminacy (or, equivalently, to fix 
$f(x)$), one has to introduce an ``external and largely arbitrary'' relation among the fields (see (\ref{Sec1:condition00})).
%If, for instance, such a relation (algebraic or differential) is %linear, then the construction of explicit solutions via 
%the dressing algorithm remains  
%linear as well. The simplest example of linear relation is %obtained imposing that one of the fields be a given function of %the 
%coordinates, like in (\ref{Sec1:relation_proto}), interpretable %as an external forcing.  
\item
 The system of PDEs depends 
on two types of matrix fields, those obtained ``saturating the parameter $\lambda$'' of the solution $U(\lambda;x)$ 
of the linear integral equation by 
ingredients of the classical dressing method, whose dimensionality is constrained, and those 
obtained saturating $\lambda$ by a novel dressing function $G(\lambda;x)$, whose dimensionality is not constrained. That's why 
the dimensionality of the solution space, ($n-2$), can be arbitrarily large. 
\item
 While integrable PDEs in low dimensions (2+1 or less) are the compatibility of overdetermined systems of  
linear spectral problems, such a feature seems to be lost for our higher dimensional examples. 
\end{enumerate}

A prototype example of the above construction is given by the following 4 dimensional system of two matrix equations 
\begin{eqnarray}\label{Sec1:equation_proto}
 {\cal B}_2(q_1,q_1,q_2){\cal B}^{-1}_2(q_1,q_2,q_3)={\cal B}_3(q_1,q_1,q_2){\cal B}^{-1}_3(q_1,q_2,q_3)=
{\cal B}_4(q_1,q_1,q_2){\cal B}^{-1}_4(q_1,q_2,q_3)
 \end{eqnarray}
for the three square matrix fields $q_1(x),q_2(x),q_3(x)$, supplemented by the ``largely arbitrary'' relation
\begin{eqnarray}\label{Sec1:condition_proto}
F(q_1,q_2,q_3) = 0
\end{eqnarray}
among them, where the matrix blocks ${\cal B}_j$ are defined as: 
\begin{equation}\label{Sec1:def_proto}
{\cal B}_j(q_1,q_2,q_3)\equiv {q_2}_{x_j}-{q_2}_{x_1}B_j-q_2[q_1,B_j]-[B_j,q_3],\;\;\;j=2,3,4,
\end{equation}
and $B_j,\;j=2,3,4$ are constant diagonal matrices different from the identity. 
In the simplest case, the largely arbitrary relation 
(\ref{Sec1:condition_proto}) can be chosen to be an equation defining one of the fields, say $q_3$, to be any given 
function $\gamma(x)$ (in general, a generalized function), interpretable as an ``external arbitrary forcing'': 
\beq
\label{Sec1:relation_proto}
F:\;\;\;\;\;q_3(x)=\gamma(x).
\eeq
The partially integrable nonlinear PDEs (\ref{Sec1:equation_proto}-\ref{Sec1:relation_proto}) possess a manifold of analytic 
solutions of dimension $2$.

%%%%%%%%%%%%%%%%%%%%%%%%%%%%%%
\subsection{Dressing method based on homogeneous Integral equation}
\label{Novelties}

Although the starting integral equation used in \cite{ZS,Z3} is still inhomogeneous, the use in \cite{Z3} 
of first order multidimensional dressing operators allows one to enrich the space of analytic solutions of 
the constructed nonlinear PDEs (although full integrability is not achieved even there), and to simplify the structure 
of such nonlinear PDEs (which become differential polynomials), if compared to the block structure of equations like 
(\ref{Sec1:equation_proto}-\ref{Sec1:relation_proto}). 

In addition, the introduction of such first order dressing operators 
makes clear that the inhomogeneous term $\Phi$ of the integral equation (\ref{Sec1:U}) is not necessary anymore, 
suggesting the new scenario, discussed in this paper, of a dressing algorithm based on the following 
{\it homogeneous} integral equation:
\begin{eqnarray}\label{U}\label{basic}
0&=&
\int \Psi(\lambda,\nu;x)
U(\nu;x) d\Omega(\nu) \equiv \Psi(\lambda,\nu;x) *
U(\nu;x),
\end{eqnarray}
 supplemented by the generalized commutation relation

\begin{eqnarray}\label{comm0}
{\cal{A}}(\lambda,\nu)* \Psi(\nu,\mu;x) = \Psi(\lambda,\nu;x)*
 A(\nu,\mu).
\end{eqnarray}

In the integral equation (\ref{U}), $U$ is the unknown  spectral function depending on the single 
spectral parameter $\lambda$; the kernel $\Psi$ of the  integral operator, the dressing 
function, satisfies the generalized commutation relation (\ref{comm0}) for a proper choice of 
the auxiliary functions ${\cal{A}}$ and $A$. Function $\Psi$ and, consequently, $U$,  depend on an 
additional set of variables 
$x=(t_1,t_2,\dots, x_1,x_2,\dots)$, which are the independent variables of the associated  
nonlinear PDEs. In this paper we assume that the functions ${\cal{A}}$ and $A$ do not depend on 
$x$.  In general, all functions are $Q\times Q$ matrices.

Following \cite{ZS}, we assume in this paper that the integral equation (\ref{U}) possesses nontrivial solutions, 
namely, that 
\begin{eqnarray}\label{d}
\dim{\mbox{ker}}(\Psi *)=d>0.
\end{eqnarray}
%Kernels of this type have been shown to be useful for 
%constructing new examples of nonlinear PDEs in 
%multidimensions having a large manifold of particular solutions 
%\cite{ZS}, \cite{Z3}.

The general solution of the homogeneous equation (\ref{U}) reads
\begin{eqnarray}\label{2_deg}
U(\lambda;x)= \sum_{i=1}^d U_h^{(i)}(\lambda;x)  f^{(i)}(x),
\end{eqnarray}
where $U_h^{(i)}$ are independent particular solutions of (\ref{U}) and  $f^{(i)}$ are 
arbitrary functions of $x$. It is convenient  to introduce a "rectangular" integral operator 
in the following way. 
Let $D$ be a set of points: $D=\{l_1,\dots,l_M\}$, and let ${\cal{D}}$ be a disjoint set 
(${\cal{D}}\cap D=\emptyset$), 
consisting eventually of continuous and discrete parts. Thus, in 
the function $\Psi(\lambda,\mu;x)$, $\lambda\in {\cal{D}}$, while $\mu \in {\cal{D}}\cup D$:
\begin{eqnarray}\label{Psi_def}
\Psi(\lambda,\mu;x) =\left\{ 
\begin{array}{ll}
\psi(\lambda,\mu;x), & \lambda,\mu\in {\cal{D}}, \cr
\psi_{0n}(\lambda;x), &\lambda\in{\cal{D}}, \;\;\mu=l_n .
\end{array}\right.
\end{eqnarray}
As a consequence of this assumption, $\lambda \in {\cal{D}}\cup D$ in $U(\lambda;x)$:
\begin{eqnarray}\label{U_def}
U(\lambda;x) =\left\{ 
\begin{array}{ll}
u(\lambda;x), & \lambda\in {\cal{D}}, \cr
u_n(x) & \lambda=l_n ,
\end{array}\right.
\end{eqnarray}
and the integral equation (\ref{U}) reduces to the form
\begin{eqnarray}\label{u}
\psi(\lambda,\mu;x)* u(\mu;x) +\sum_{i=1}^M \psi_{0i}(\lambda;x) u_i(x) = 0,~~~
\lambda \in {\cal{D}}.
\end{eqnarray}
If the integral operator $\psi(\lambda,\mu;x)*$ is invertible, the solution 
$u(\lambda;x)$ is uniquely expressed in terms of the $M$ arbitrary functions $u_i(x)$, 
which may be 
identified with the functions $f^{(i)}(x)$ in eq.(\ref{2_deg}), and $d=M$. If $\psi(\lambda,\mu;x)*$ 
is not invertible, then $d>M$.
 As in \cite{ZS}, we introduce an external dressing function $G(\lambda;x)$ in the next section in order to fix $f^{(i)}(x)$. 

We remark that the integral equation (\ref{u}) can be viewed as an inhomogeneous 
integral equation  
with an inhomogeneous term depending on $M$ arbitrary functions. In the rest of the paper 
we find it more  convenient to work with the homogeneous form (\ref{U}).

The rectangular  structure of $\Psi$ implies also different "square" structures 
for the integral operators ${\cal{A}}*$ and $*A$ in (\ref{comm0}):
\begin{eqnarray}\label{cA_def}
{\cal{A}}(\lambda,\mu) =\left\{ 
\begin{array}{ll}
{\sl{a}}(\lambda,\mu), & \lambda,\mu\in {\cal{D}} \cr
0, &\lambda\in D \;\;{\mbox{or}} \;\;\mu\in D,
\end{array}\right.
\end{eqnarray}
\begin{eqnarray}\label{A_def}
A(\lambda,\mu) =\left\{ 
\begin{array}{ll}
a(\lambda,\mu), & \lambda,\mu\in {\cal{D}} \cr
a_{0m}(\lambda), &\lambda\in{\cal{D}}, \;\;\mu=l_m \cr
a_{n0}(\mu), &\lambda=l_n, \;\;\mu \in {\cal{D}}\cr
a_{nm}, & \lambda=l_n,\;\;\mu=l_m
\end{array}\right.,
\end{eqnarray}
where $n,m=1,\dots,M$. 

To end this section, let us carry out a comparison of algorithms developed in  \cite{ZS} and in this paper, which we call Alg.1 and Alg.2 respectively.
\begin{enumerate}
\item
Both  algorithms use the nontrivial kernel of the integral operators, but the integral equation is inhomogeneous  in Alg.1 and homogeneous in Alg.2.
\item
Both algorithms use two types of   
dressing functions: external and internal dressing functions. 
However, Alg.1 uses two internal dressing functions, $\Phi(\lambda,\mu)$ and $C(\mu;x)$ with the kernel $\Psi$ expressed through $\Phi$ and $C$, while Alg.2 uses the single internal dressing function $\Psi(\lambda,\mu;)$, 
the kernel of the integral operator. 
\item
Both algorithms use two disjoint domains on the spectral parameter plane: 
a continuous ${\cal{D}}$ and a discrete $D$. But the number of points $M$ in $D$ is minimized in Alg.2, 
while this number is an arbitrary $M>\dim{\mbox{ker}} \hat \Psi$ in Alg.1.
\item
Alg.2 describes the full solution space of the appropriate $n$-dimensional nonlinear PDEs, while Alg.1 describes  
only an $(n-2)$ dimensional subspace of analytic solutions.
\item
Both algorithms deal with nonlinear PDEs which may not be considered as necessary compatibility condition of some overdetermined linear system of PDEs for the associated spectral function.
\item
Both systems of PDEs derived by Alg.1. and Alg.2 have infinitely many commuting flows.
\end{enumerate}

In this paper we concentrate on the case $M=1$.   
Before giving more  details regarding the  solvability of eqs.(\ref{U}) and (\ref{comm0}), 
we present the derivation of eq.(\ref{char}).

%%%%%%%%%%%%%%%%%%%%%%
\section{First order quasi-linear PDEs}
\label{First_order}
%%%%%%%%%%%%%%%%%%%%%
\subsection{Derivation of the PDEs (\ref{char})}
As usual in the dressing philosophy, the 
$x$-dependence of the spectral function $U(\lambda;x)$ is introduced through the $x$-dependence of 
the dressing functions. In this paper, as well as in the dressing algorithm introduced 
in \cite{ZS},  we have two types of dressing functions. The internal dressing function 
$\Psi(\lambda,\mu;x)$, appearing in the integral equation (\ref{U}), and an external 
dressing function  $G(\lambda;x)$, $\lambda\in {\cal D}\cup D$:
 
\begin{eqnarray}\label{G_def}
G(\lambda;x) =\left\{ 
\begin{array}{ll}
g(\lambda;x), & \lambda\in {\cal{D}}, \cr
g_1(x) & \lambda = l_1 ,
\end{array}\right.
\end{eqnarray}
whose role will be explained below. 

This $x$-dependence is defined by the equations
\begin{eqnarray}\label{x}
&&
\Psi_t(\lambda,\mu;x)+
\sum_{j=1}^{N} {\cal{A}}^{(j)}(\lambda,\nu)*\Psi_{x_j}(\nu,\mu;x) = 0,\\\label{G}
&&
G_t(\lambda;x)+
\sum_{j=1}^{N}G_{x_j}(\nu;x)*A^{(j)}(\nu,\lambda)
 = -T G(\nu;x) *A^{(0)}(\nu;\lambda) ,
\end{eqnarray}
supplemented by the generalized commutation relation:
\begin{eqnarray}\label{commj}
{\cal{A}}^{(j)}(\lambda,\nu)* \Psi(\nu,\mu;x) = \Psi(\lambda,\nu;x)*
 A^{(j)}(\nu,\mu),
\end{eqnarray}
where $T$ is an arbitrary constant  matrix. Since 
${\cal{A}}^{(j)}*$ and $*A^{(j)}$ satisfy the same relation as ${\cal{A}}*$ and $*A$, 
it follows that 
they are expressed as functions of  the operators ${\cal{A}}*$ and $*A$ respectively:
\begin{eqnarray}\label{rho_def_1}
{\cal{A}}^{(j)}*\equiv \rho^{(j)}({\cal{A}})*,~~~~*A^{(j)}\equiv *\rho^{(j)}(A), 
\end{eqnarray}
where $\rho^{(j)}(\cdot)$ are scalar functions representable as positive power series:
\begin{eqnarray}\label{rho_def_2}
\rho^{(j)}(y)=\sum\limits_{k=0}^{\infty}c^{(j)}_ky^k~~\Rightarrow~~\left\{
\ba{c}
*A^{(j)}=
*\rho^{(j)}(A)=\sum\limits_{k=0}^{\infty}c^{(j)}_k*\underbrace{A*\dots *A}_k \\
{\cal A}^{(j)}*=\rho^{(j)}(A)*=\sum\limits_{k=0}^
{\infty}c^{(j)}_k\underbrace{{\cal A}*\dots *{\cal A}}_k*
\ea
\right. 
\end{eqnarray}
From the above definitions and from (\ref{A_def}) it follows that 
\begin{eqnarray}\label{A_def_j}
A^{(j)}(\lambda,\mu) =\left\{ 
\begin{array}{ll}
a^{(j)}(\lambda,\mu), & \lambda,\mu\in {\cal{D}} , \cr
a^{(j)}_{01}(\lambda), &\lambda\in{\cal{D}}, \;\;\mu=l_1 , \cr
a^{(j)}_{10}(\mu), &\lambda=l_1, \;\;\mu \in {\cal{D}} , \cr
a^{(j)}_{11}, & \lambda=l_1,\;\;\mu=l_1 .
\end{array}\right.
\end{eqnarray}

Applying the operator ${\cal{A}}*$ to  
eq.(\ref{U}) and using (\ref{comm0}), one shows that $A*U$ is another solution of the integral 
equation (\ref{U}), i.e.:
\begin{eqnarray}
\Psi * (A*U) =0.
\end{eqnarray}

Applying instead the operator ($\partial_t + \sum_{i=1}^N {\cal{A}}^{(i)} \partial_{x_i} *$)  
to eq.(\ref{U}) 
and using the generalized commutation relation (\ref{comm0})
and equation (\ref{x}), one obtains a third solution of the integral equation (\ref{U}):
\begin{eqnarray}
\Psi(\lambda,\mu;x) *
 \hat LU(\mu;x) =0 ,\;\;\;\hat LU(\mu;x)\equiv
U_t(\mu;x) + 
\sum_{j=1}^N A^{(j)}(\mu,\nu)* U_{x_j}(\nu;x).
\end{eqnarray}

In this section we assume that the three solutions $U$, $A*U$ and $\hat LU$ of the integral 
equation (\ref{U}) belong to the same one dimensional matrix subspace spanned by $U$; i.e.:
\begin{eqnarray}\label{3sp_2}
A(\lambda,\nu)*U(\nu;x) = U(\lambda;x) \tilde F(x)
\end{eqnarray}
\begin{eqnarray}\label{3sp_1}
U_t(\lambda;x) + 
\sum_{j=1}^N A^{(j)}(\lambda,\nu)* U_{x_j}(\nu;x)=
U(\lambda;x) F(x),
\end{eqnarray}
where the matrices $\tilde F$ and $F$ do not depend on the spectral parameters. 

As in the dressing scheme introduced in \cite{ZS}, in order to fix $\tilde F$ and $F$, 
we use the external dressing function $G$, together with the additional relation:
\begin{eqnarray}\label{condition}
G(\lambda;x)*U(\lambda;x)=I,
\end{eqnarray}
where $I$ is the identity matrix. 

Applying $G*$ to equation (\ref{3sp_2}) and using (\ref{condition}), 
one obtains that
\begin{eqnarray}\label{tilde_F}
\tilde F(x)=w(x),
\end{eqnarray}
where 
\begin{eqnarray}\label{def_w}
w(x)\equiv G(\lambda;x)*A(\lambda,\mu)*U(\mu;x);
\end{eqnarray}
so that
\begin{eqnarray}\label{AU=Uw}
A(\lambda,\nu)*U(\nu;x) = U(\lambda;x) w(x) .
\end{eqnarray}
In addition, applying repeatedly $A*$ to equation (\ref{AU=Uw}),  we obtain 
\begin{eqnarray}\label{AiU}
\rho(A)*U=U\rho(w),
\end{eqnarray}
where $\rho:\;\RR\to\RR$ is any scalar analytic function. 
Therefore equation (\ref{3sp_1}) becomes:
\begin{eqnarray}\label{3sp_1_bis}
\ba{l}
U_t(\lambda;x) + \sum_{j=1}^N \left(U(\lambda;x)\rho^{(j)}(w)\right)_{x_j}=
U(\lambda;x) F(x).
\ea
\end{eqnarray}
Applying now $G*$ to (\ref{3sp_1_bis}) and using (\ref{condition}) and (\ref{AiU}), 
one obtains
\begin{eqnarray}\label{def_F}
F(x)=\sum_{j=1}^N\left(\rho^{(j)}(w)\right)_{x_j}+ T\rho^{(0)}(w);
\end{eqnarray}
so that the system system (\ref{3sp_2}), (\ref{3sp_1}) takes the final form:
\begin{eqnarray}\label{Usystem}
\ba{l}
A(\lambda,\nu)*U(\nu;x) = U(\lambda;x) w(x) , \\
U_t(\lambda;x) + \sum_{j=1}^N U_{x_j}(\lambda;x)\rho^{(j)}(w)=
U(\lambda;x)T\rho^{(0)}(w) .
\ea
\end{eqnarray}
It is a system of overdetermined linear equations 
for the spectral function $U$ but, as we shall see in Sec. 6, its role is different from that played by the 
usual Lax pair for soliton equations.   
The main feature of the linear equation (\ref{Usystem}a) is that it does not involve $x$-derivatives. 

Applying $G*A*$ to (\ref{Usystem}b), one finally obtains the matrix equation  
\begin{eqnarray}\label{3nl_1}
w_{t} + 
\sum_{j=1}^N  w_{x_j} \rho^{(j)}(w)=[w, T\rho^{(0)}(w)]
\end{eqnarray}
reported in the introduction as equation (\ref{char}). We recall that $w$ may be either a 
scalar or a matrix. 

%%%%%%%%%%%%%%%%%%%%%%%%%%
\subsection{Basic properties of eq.(\ref{3nl_1})}
\label{Reductions}
The matrix first order quasilinear PDE (\ref{3nl_1}), isolated by the above dressing construction, 
possesses important properties and can be integrated by simple spectral 
means \cite{SZ}. 
Here we summarize, for completeness, some of these properties.
   
\vskip 5pt
\noindent
1.  We first observe that, if $w$ solves eq.(\ref{3nl_1}), then 
$w^{T}$ solves the transposed equation
\begin{eqnarray}\label{3nl_1tr}
w^T_{t} + 
\rho^{(j)}(w^T)\sum_{j=1}^N  w^T_{x_j} =[\rho^{(0)}(w^T) T^T,w^T].
\end{eqnarray}
where the order of matrix multiplication is inverted. We also observe that $T$ can be chosen to be diagonal 
without loss of generality; indeed, if it were not diagonal, the diagonalizing similarity transformation 
for $T$ would leave eq.(\ref{3nl_1}) invariant, just transforming $w$ according to the same  
similarity transformation.

\vskip 5pt
\noindent
2. Let $T$ be the diagonal matrix, $e^{(i)}$ and $\underline{v}^{(i)}$, $i=1,\dots,Q$ be the eigenvalues and the right 
eigenvectors of the matrix $w$, suitably 
normalized 
by the conditions ${v}^{(i)}_i=1,~i=1,\dots,Q$; introduce the matrix $E$ of the eigenvalues: $E=$ diag 
$(e^{(1)},\dots,e^{(Q)})$ , and the matrix $V$ having the eigenvectors $\underline{v}^{(i)}$ 
as columns ($V_{ij}={v}^{(j)}_i$). Then, if 
$w$ evolves according to eq.(\ref{3nl_1})
%, in the case in which $T$ is diagonal 
%($T_{ij}=T_i\delta_{ij}$), 
the spectrum of $w$ evolves in the following simple way:
\begin{eqnarray}\label{ee}
&&
e^{(k)}_t + \sum_{j=1}^N e^{(k)}_{x_j} \rho^{(j)}(e^{(k)}) =0,\;\;k=1,\dots,Q,\\
\label{VV}
&&
V_t+ \sum_{j=1}^N V_{x_j} \rho^{(j)}(E) =[V,T]\rho^{(0)}(E).
\end{eqnarray}

It follows from (\ref{ee}) that each eigenvalue of $w$ evolves separately according to the scalar 
version of (\ref{3nl_1}), and is constant along the characteristic straight lines 
\beq \label{j-curve}
x_j=\rho^{(j)}(e^{(k)})t+\eta_j,~~j=1,..,n, 
\eeq 
and $\eta_j$ are arbitrary integration constants. Therefore it is defined by the implicit equation 
\beq
\label{e-sol} e^{(k)}=\epsilon^{(k)}\left(\eta_1,..,\eta_N\right)=
\epsilon^{(j)}\left(x_1-\rho^{(1)}(e^{(k)})t,..,x_N-\rho^{(N)}(e^{(k)})t\right),
\eeq 
where $\epsilon^{(k)}:\RR^n\to\RR,~~k=1,\dots,Q$ are arbitrary function. 
Once the eigenvalues are constructed, then $V$ satisfies the linear equation (\ref{VV}), 
with coefficients depending on the corresponding eigenvalues. 

\vskip 5pt
\noindent
3. {\it First reduction}. From equation (\ref{ee}) it follows that the condition
\begin{eqnarray}\label{red1}
e^{(i)}(x)=\mbox{const},~~~i=1,\dots,Q,
\end{eqnarray}
represents a symmetry reduction of the system (\ref{3nl_1}). 
In this case, eq.(\ref{VV}) becomes a linear system of $(Q^2-Q)$ equations with constant 
coefficients; therefore the transformation from $w$ to its spectrum linearizes the flow.  

\vskip 5pt
\noindent
4. {\it Second reduction}. 
Consider the subspace of $Q\times Q$ matrices spanned by the basis
$\{\omega_0,..,\omega_{Q-1}\}$ given by:
\beq \label{red2}
\omega_0=I,~\left(\omega_1\right)_{ij}=\delta_{i+1,j},~..~,~
\left(\omega_k\right)_{ij}=\delta_{i+k,j},~..~,
~\left(\omega_{N-1}\right)_{ij}=\delta_{i+N-1,j},~~~\mbox{mod} Q.
\eeq
This subspace is left invariant under matrix multiplication,
since: %
\beq
\omega_j\omega_k=\omega_k\omega_j=\omega_{j+k},~~~ \mbox{mod} Q;
\eeq
therefore it defines a reduction of (\ref{3nl_1}) from the %
$Q^2$ components of $w$ to the $Q$ scalar coefficients
$\nu_k,~k=1,..,Q$ of the expansion: 
\beq \label{reduction}
w=\sum\limits_{k=1}^{Q}\nu_k\omega_{k-1}. 
\eeq
In particular, if $Q=2,~N=1$, $\rho^{(1)}(y)=y$ and $\rho^{(0)}=0$, equation 
(\ref{3nl_1}) reduces to 
the following particular case of the gas dynamics equations \cite{Whitham}
\beq \label{gas2}
\ba{l}
{\nu_1}_t+\nu_1{\nu_1}_{x_1}+\nu_2{\nu_2}_{x_1}=0, \\
{\nu_2}_t+\nu_2{\nu_1}_{x_1}+\nu_1{\nu_2}_{x_1}=0.
\ea
\eeq

\vskip 5pt
\noindent
5. The matrix equation (\ref{3nl_1tr}) can be written as a vector system of the following  form
\begin{eqnarray}
{\bf w}_t + \sum_{i=1}^N C^{(i)}({\bf w}) {\bf w}_{x_i} = {\bf B}({\bf w}) 
\end{eqnarray}
for the $Q^2$-dimensional vector {\bf w}, where
\begin{eqnarray}\nonumber 
&&
{\bf w} = \left[\begin{array}{c}
{\bf w}_1 \cr \vdots \cr
{\bf w}_Q
\end{array}
\right],
w=({\bf w}_1 \cdots  {\bf w}_Q),\;\;{\bf w}_i=\left[
\begin{array}{c}
w_{1i}\cr
\vdots\cr
w_{Qi}
\end{array}
\right],\;\;
C^{(i)}({\bf w})=\left[
\begin{array}{ccc}
\rho^{(i)}_{11} I &\cdots& \rho^{(i)}_{Q1} I \cr
 \vdots &\vdots &\vdots \cr
 \rho^{(i)}_{1Q} I &\cdots& \rho^{(i)}_{QQ} I
\end{array}\right],\\\nonumber
&&
{\bf B}({\bf w}) =
\left[\begin{array}{c}
{\bf B}_1({\bf w}) \cr \vdots \cr
{\bf B}_Q({\bf w})
\end{array}
\right],\;\;
{\bf B}_ k({\bf w}) =
\left[\begin{array}{c}
[\rho^{(n)}(w^T) T^T,w^T]_{k1} \cr\vdots \cr
[\rho^{(n)}(w^T) T^T,w^T]_{kQ}\end{array}
\right],
\end{eqnarray} 
and $I$ is the $Q\times Q$ identity matrix. By construction, the $e^{(j)}$'s are 
eigenvalues of all matrices $C^{(i)}$ with  multiplicity $Q$. Thus the method of integration 
of vector first order quasi-linear equations developed in \cite{ts1,dn,ts2} may not be applied, 
at least in the form presented in these references.
 
Hereafter $T$ is {\bf diagonal} constant matrix.
   
%%%%%%%%%%%%%%%%%%%%%%%%%%

\section{The general solution of eq.(\ref{3nl_1})}
\label{Solutions}
%%%%%%%%%%%%%%%%%
In this section we construct, using the dressing method introduced in this paper, 
the general solution of the matrix equation (\ref{3nl_1}), which 
turns out to be characterized by a nonlinear system of non-differential equations for the 
components of matrix $w$ in the following way.

\vskip 5pt
\noindent
{\bf Proposition 1}. Let $F_{ij}:\RR^N\to\RR,\;i,j=1,\dots,Q$ be $Q^2$ arbitrary scalar functions, 
representable by positive power series, so that $F_{ij}(M_1,\dots,M_N)$ are well defined matrix functions, 
where $M_1,\dots,M_N$ are arbitrary $Q\times Q$ matrices. Let $\{T_1,\dots,T_Q\}$ be the elements 
of the constant diagonal matrix $T$. Then the general solution of the 
matrix PDE (\ref{3nl_1}) is characterized by the following system of $2Q^2$ non-differential equations: 
\beq\label{sol1_w}
w_{\alpha\beta} = \sum\limits_{\delta=1}^Q \sum\limits_{\gamma_1,\gamma_2=1}^Q((u_1(x))^{-1})_{\alpha\delta}
 \Big(F_{\delta\gamma_1}(x_1 I - \rho^{(1)}(w) t,\dots,x_N I - \rho^{(N)}(w) t)\Big)_{\gamma_2\beta} 
(u_1(x))_{\gamma_1\gamma_2},
\eeq
\begin{eqnarray}\label{sol1_u1}
\sum\limits_{\gamma=1}^Q (u_1(x))_{\alpha\gamma}
\Big(e^{-\rho^{(0)}(w)T_{\alpha}t}\Big)_{\gamma\beta}= \delta_{\alpha\beta},\;\;\alpha,\beta=1,\dots,Q.
\end{eqnarray}
for the components of the matrix solution $w(x)$ and of the auxiliary matrix function $u_1(x)$. 

\vskip 5pt
\noindent
{\bf Remarks} \\
{\bf 1}. If $T=0$, equation (\ref{sol1_u1}) gives $u_1=I$; 
then equation  (\ref{sol1_w}) reduces to the following system of $Q^2$ non-differential equations for the 
components of $w$: 
\begin{eqnarray}\label{sol3}
\label{w_f}
w_{\alpha\beta} = \sum_{\gamma=1}^Q 
\Big(F_{\alpha\gamma}(x_1 I - \rho^{(1)}(w) t,\dots,x_N I - \rho^{(N)}(w) t)\Big)_{\gamma\beta} ,\;\;
\alpha,\beta=1,\dots,Q.
\end{eqnarray}
This equation can also be constructed using equations (\ref{ee}) and (\ref{VV}) for $T=0$ \cite{SZ}. \\
{\bf 3}. If, at last, one is interested in the scalar version of equation (\ref{3nl_1}), 
its general solution, characterized by the scalar version of (\ref{sol3}):
\beq\label{sol_scalar}
w=F\Big(x_1- \rho^{(1)}(w) t,\dots,x_N - \rho^{(N)}(w) t\Big) ,
\eeq
is also easily obtainable using the method of characteristics. \\
{\bf 4}. Due to the presence of the $Q^2$ arbitrary scalar functions $F_{ij},\;i,j=1\dots,Q$, 
the above non-differential equations characterize the general solution of the matrix PDE (\ref{3nl_1}). 
In particular, if one is interested in solving the Cauchy problem in $\RR^N$ with the prescribed initial 
condition $w_0(x_1,\dots,x_N)\equiv w(x_1,\dots,x_N,0)$,  equation (\ref{sol1_u1}), evaluated at $t=0$, 
implies that $u_1|_{t=0}=I$. Then equation (\ref{sol1_w}) at $t=0$ implies that 
\beq
F(x_1,\dots,x_N)=w_0(x_1,\dots,x_N).    
\eeq
Known the functions $F_{ij}$, 
the nonlinear non-differential equations
(\ref{sol1_w},\ref{sol1_u1}) allow one to 
construct, $\forall t$, the solution $w(x_1,\dots,x_N,t)$.
 
%%%%%%%%%%%%%%%%%%%%%%%%%%%%%%%%%%%%%
\subsection{Proof of Proposition 1}
To prove Proposition 1, we make use of the main ingredients of the dressing scheme:   
eqs.(\ref{3sp_2}),(\ref{tilde_F}) and 
(\ref{condition}), written explicitly using (\ref{U_def},\ref{A_def},\ref{G_def}):
\begin{eqnarray}\label{SPU}
&&
a(\lambda,\mu)*u(\mu;x) + a_{01}(\lambda) u_1(x)= u(\lambda;x) w(x),\;\;
\lambda\in {\cal{D}},\\\nonumber
&&
a_{10}(\mu)*u(\mu;x) + a_{11} u_1(x)= u_1(x) w(x),\;\;\lambda=l_1 ,
\\\label{SPG}\label{condition_sol}
&&
g(\mu;x)*u(\mu;x) + g_1(x) u_1(x) = I ,
\end{eqnarray}
eq.(\ref{def_w}) for $w$:
\begin{eqnarray}\label{w_mod}\label{w_g}\label{w_dr} 
w(x)&\equiv& G(\lambda;x)*A(\lambda,\mu)*U(\mu;x)=
g(\lambda;x)*a(\lambda,\mu) *u(\mu;x) + \\\nonumber
&&
g(\lambda;x)*a_{01}(\lambda) u_1(x) +
g_1(x) \Big(a_{10}(\lambda)*u(\lambda;x)+a_{11} u_1(x)\Big) ,
\end{eqnarray}
and the generalized 
commutation relation (\ref{comm0}), which we rewrite to emphasize the fact that  the parameters
$\lambda$ and $\mu$ take values in different domains:
\begin{eqnarray}\label{comm}
{\cal{A}}(\lambda,\nu)* \Psi(\nu,\mu) = \Psi(\lambda,\nu)*
 A(\nu,\mu), \;\;\;\;\;\;\lambda \in {\cal{D}},\;\; \mu \in {\cal{D}} \cup D.
\end{eqnarray}

Using formulae (\ref{Psi_def}), (\ref{U_def}), (\ref{cA_def}) and (\ref{A_def}),
eq.(\ref{comm}) takes the form:
\begin{eqnarray}\label{comm_sol0}
&&
{\sl{a}}(\lambda,\nu)* \psi(\nu,\mu;x)=\psi(\lambda,\nu;x)* a(\nu,\mu) +
\psi_{01}(\lambda;x) a_{10}(\mu),\;\;\lambda,\mu \in {\cal{D}},\\\label{comm_sol20}
&&
{\sl{a}}(\lambda,\nu)* \psi_{01}(\nu;x)=\psi(\lambda,\nu;x)* a_{01}(\nu) +
\psi_{01}(\lambda;x) a_{11},\;\;\lambda\in {\cal{D}},~\mu =l_1.
\end{eqnarray} 
This system should be viewed as a system of linear equations for the functions
$\psi$ and $\psi_{01}$, where ${\sl a}$, $a$, $a_{ij}$ are largely arbitrary.
The following two  choices for ${\sl a}$ and $a$ have been explored so far:
\begin{eqnarray}\label{a_ex1}
1.&&\hspace{3cm}{\sl a}(\lambda,\mu) = a(\lambda) \delta(\lambda-\mu) ,\;\;
{a}(\lambda,\mu) = a(\lambda) \delta(\lambda-\mu),\\
\label{a_ex2}
2.&&\hspace{3cm}{\sl a}(\lambda,\mu) =  i\delta'(\lambda-\mu) ,\;\;
{a}(\lambda,\mu) = -i\delta'(\lambda-\mu),
\end{eqnarray}
where $\delta$ and $\delta'$ are the Dirac function and its derivative. 

It is possible to verify that the first choice (\ref{a_ex1}) corresponds to the 
case in which the dressing function $\Psi$ does not depend on $x$ and, consequently, 
the eigenvalues of $w$ are constant as well. Therefore the choice (\ref{a_ex1}) 
leads to the trivial reduction discussed in the previous section (see (\ref{red1}). 

As we shall see in the following, the second choice (\ref{a_ex2}) allows one to capture instead the 
general solution of the matrix PDE (\ref{3nl_1}).

%%%%%%%%%%%%%%%%%%%%%%%%%
\subsubsection{Solution space associated with eqs.(\ref{a_ex2})}
\label{Psix_neq_0}
\label{Basic_eq}

%%%%%%%%%%%%%%%%%%%%%%%%%%%%
\paragraph{Construction of the dressing function $\Psi(\lambda,\mu;x)$}

Under the assumption (\ref{a_ex2}), function 
 $\Psi$ is completely defined by the equation (\ref{x}),
which reads
\begin{eqnarray}\label{ex2:t}
&&\psi_t(\lambda,\mu;x) +\sum_{j=1}^N \rho^{(j)}(i\partial_\lambda) \psi_{x_j} (\lambda,\mu;x) 
=0 ,\;\;
\lambda,\mu \in {\cal{D}},\\
\label{ex2:t2}
&&{\psi_{01}}_t(\lambda;x) +\sum_{j=1}^N \rho^{(j)}(i\partial_\lambda) {\psi_{01}}_{x_j} (\lambda;x) 
=0,\;\; \lambda\in {\cal{D}},\;\mu=l_1,
\end{eqnarray}
and by the system  (\ref{comm_sol0},\ref{comm_sol20}):
\begin{eqnarray}\label{ex2:comm}
&&
i \psi_{\lambda}(\lambda,\mu;x) = 
i\psi_{\mu} (\lambda,\mu;x)  + 
\psi_{01}(\lambda;x) a_{10}(\mu),\;\;\lambda,\mu \in {\cal{D}},\\
\label{ex2:comm2}
&&
i{\psi_{01}}_{\lambda}(\lambda;x) = 
\psi (\lambda,\nu;x)*a_{01}(\nu)  + \psi_{01}(\lambda;x) a_{11},\;\;\lambda\in {\cal{D}},\;\mu = l_1.
\end{eqnarray}

Equations (\ref{ex2:t}-\ref{ex2:comm2}) suggest  
to represent $\psi$, $\psi_{01}$ and  $a_{10}$  in the Fourier forms
\begin{eqnarray}\label{Fourier}
&&\psi(\lambda,\mu;x)=\int\limits_{-\infty}^\infty 
d\varkappa \int\limits_{-\infty}^\infty d q \int\limits_{\RR^N} dk 
\;\tilde \psi(\varkappa,q,k) 
e^{i \varkappa \lambda + i q \mu + i \sum\limits_{j=1}^N k_j [x_j -
 \rho^{(j)}(-\varkappa) t]},\\
&&\psi_{01}(\lambda;x)=\int\limits_{-\infty}^\infty  
d\varkappa\int\limits_{\RR^N}  dk 
\;\tilde \psi_{01}(\varkappa,k) 
e^{i \varkappa \lambda  + i  \sum\limits_{j=1}^N k_j [x_j -
 \rho^{(j)}(-\varkappa) t]} ,\\\label{a1001}
&&a_{10}(\mu)= \int\limits_{-\infty}^\infty\tilde a_{10}(q) 
e^{i q \mu } dq,\\\label{u_F}
&&
u(\lambda;x)=\int\limits_{-\infty}^\infty \frac{dq}{2\pi}\;
\tilde u(q;x) e^{i q\lambda},
\end{eqnarray}
where $k=(k_1,\dots,k_N)$.

For future convenience, we also represent $a_{01}(\mu)$ via the contour integral 
\begin{eqnarray}\label{2_a01}
a_{01}(\mu)=\frac{1}{2\pi i}\oint\limits_{\Gamma_{01}} \tilde a_{01}(q) 
e^{ iq \mu } dq,
 \end{eqnarray}
where $\Gamma_{01}$ is  a sufficiently large contour containing all singularities of the integrand.

With the above representations, eqs.(\ref{ex2:t},\ref{ex2:t2}) are automatically satisfied, while
eqs.(\ref{ex2:comm},\ref{ex2:comm2}) yield the relations:
\begin{eqnarray}\label{ex2:comm21}
&&
\tilde \psi(\varkappa,q,k) (\varkappa-q) = -\tilde \psi_{01}(\varkappa,k) \tilde a_{10}(q),
\\\label{ex2:comm22}
&&
\tilde \psi_{01}(\varkappa,k) ( \varkappa+a_{11}) = 
- \int\limits_{-\infty}^\infty d\mu \int\limits_{-\infty}^\infty dq\;
\tilde \psi(\varkappa,q,k) e^{iq \mu} a_{01}(\mu).
\end{eqnarray}
We solve the first of these equations with respect to  $\tilde \psi$
\begin{eqnarray}\label{psi_F}
\tilde \psi(\varkappa,q,k)  = -
\frac{\tilde \psi_{01}(\varkappa,k) \tilde a_{10}(q)}{\varkappa-q} +
 \phi(\varkappa,k)\delta(\varkappa-q),
\end{eqnarray}
where $\phi$ is an arbitrary function of its arguments, and
substitute the result in (\ref{ex2:comm22}):
\begin{eqnarray}\label{psi01}
&&\tilde \psi_{01}(\varkappa,k) \Big(\varkappa+a_{11}-\eta(\varkappa)\Big) = 
-\int\limits_{-\infty}^\infty d\mu \;
\phi(\varkappa,k)e^{i \varkappa \mu}
a_{01}(\mu) ,\;\;\\\nonumber
&&\hspace{3cm}
\eta(\varkappa) = \int\limits_{-\infty}^\infty
 d\mu\int\limits_{-\infty}^\infty dq\;  
 \frac{\tilde a_{10}(q) e^{i q \mu} 
 a_{01} (\mu)}{\varkappa-q} 
,
\end{eqnarray}
We take the following solution of this equation:
\begin{eqnarray}\label{varphi}
\ba{l}
\tilde \psi_{01}(\varkappa,k)=-
\phi(\varkappa,k) \int\limits_{-\infty}^\infty d\mu \;
e^{i \varkappa \mu}a_{01}(\mu) 
\Omega^{-1}(\varkappa), \\
 \Omega(\varkappa) \equiv  \varkappa+a_{11}-\eta(\varkappa);
\ea
\end{eqnarray}
then eq.(\ref{psi_F}) yields
\begin{eqnarray}\label{psi_tilde}
\tilde \psi(\varkappa,q,k) = \phi(\varkappa,k)\left[ 
\int\limits_{-\infty}^\infty d\mu \; 
\frac{e^{i\varkappa \mu} a_{01}(\mu) \Omega^{-1}(\varkappa) \tilde a_{10}(q) }{\varkappa - q} + 
\delta(\varkappa-q) \right].
\end{eqnarray}
Substitute $\tilde \psi$ and $\tilde \psi_{01}$ into (\ref{u}), with $M=1$,  and 
apply the operator $\frac{1}{2\pi} \int\limits_{-\infty}^\infty d \lambda\; 
 e^{i \xi \lambda} \cdot$ to  the result, obtaining:
\begin{eqnarray}\label{int}
\tilde \phi \Big(-\xi,x_1-\rho^{(1)}(\xi) t,\dots,
 x_N-\rho^{(N)}(\xi) t \Big)\times\\\nonumber
\left[\tilde u(\xi;x)+ \left(\int\limits_{-\infty}^\infty d \nu a_{01}(\nu)e^{-i\xi\nu}\right)\Omega^{-1}(-\xi)
\left(\int\limits_{-\infty}^\infty \frac{d q}{q-\xi}\; \tilde a_{10}(-q) \tilde u(q;x) 
-u_1(x)\right)\right]=0,
\end{eqnarray}
where 
\begin{eqnarray}\label{tphi2}
&&
\tilde \phi\Big(-\xi,x_1-\rho^{(1)}(\xi) t,\dots,
 x_N-\rho^{(N)}(\xi) t\Big) =\int\limits_{\RR^N} \phi(-\xi,k) 
e^{i \sum_{j=1}^N k_j(x_j-\rho^{(j)}(\xi) t)} d k.
\end{eqnarray}

We also remark that eq.(\ref{SPU}) takes the form
\begin{eqnarray}\label{u_Gam}
&&
-i u_\lambda(\lambda;x) + a_{01}(\lambda) u_1(x) = u(\lambda;x) w(x)
,\;\;\lambda \in {\cal{D}} , \\\label{w_Gam}
&&
a_{10}(\mu)*u(\mu;x) + a_{11} u_1(x) = u_1(x) w(x),\;\;\lambda=l_1.
\end{eqnarray}
Due to equation (\ref{2_a01}), the solution of (\ref{u_Gam}) leads to the following 
representation of $u$:
\begin{eqnarray}\label{uu1_mod}
u(\lambda;x)=\frac{1}{2\pi i}\oint\limits_{\Gamma_{01}}d q \;
 \tilde a_{01}(q)e^{i q \lambda}  u_1(x)\Big(w(x)-qI\Big)^{-1} , 
\end{eqnarray}
different from the Fourier representation (\ref{u_F}). 
Then eq.(\ref{w_Gam}) yields:
\begin{eqnarray}\label{condition_a}
\frac{1}{2\pi i}\int\limits_{-\infty}^\infty d\mu \oint\limits_{\Gamma_{01}}d q\; 
a_{10}(\mu)\tilde a_{01}(q) e^{i q\mu}u_1(x) \Big(w(x)- q\Big)^{-1}   = u_1(x)
 w(x) - a_{11} u_1(x) . 
\end{eqnarray}
The formulae derived so far are applicable to both the scalar and the matrix equations. 
Below we consider these two cases separately, starting with the simpler one.

%%%%%%%%%%%%%%%%%%
\paragraph{Scalar nonlinear PDEs}
\label{Scalar}
In the scalar version  
\begin{eqnarray}\label{scalar_3nl_1}
w_{t} + \sum_{j=1}^N  w_{x_j} \rho^{(j)}(w)=0
\end{eqnarray}
of (\ref{3nl_1}) the RHS is zero. In this case we can set $T=0$; therefore 
equation (\ref{G}) admits the trivial solution $g=0$, $g_1=1$;  so that $u_1=1$ (see 
(\ref{condition_sol})). 

It follows that 
 eq.(\ref{w_mod}) coincides with eq.(\ref{w_Gam}), while equation (\ref{condition_a}) reads:
 \begin{eqnarray}\label{a10a01}
\frac{1}{2\pi i}\int\limits_{-\infty}^\infty d\mu \oint\limits_{\Gamma_{01}} d q\;
 \frac{a_{10}(\mu)\tilde a_{01}(q) e^{iq\mu}}{w-q}  = 
 w - a_{11}  .
\end{eqnarray}
This equation indicates that, if $a_{10}$ and $a_{01}$ are given, then (\ref{a10a01}) is an 
algebraic constraint for $w$, which would imply that $w$ is constant. To avoid such a 
trivialization, eq.(\ref{a10a01}) must be taken as a definition  of $a_{10}$ or $a_{01}$, 
and $w$ must be considered as the independent variable.

In addition, one should make sure that the condition (\ref{a10a01}) does not coincide with 
the condition $\Omega(\varkappa)=0$ (see (\ref{varphi}b)); i.e.: 
\begin{eqnarray}\label{eta_neq}
\eta(\varkappa)\equiv  \int\limits_{-\infty}^\infty d\mu 
 \int\limits_{-\infty}^\infty dq\;  \frac{\tilde a_{10}(q) e^{iq \mu} 
 a_{01} (\mu)}{\varkappa-q} 
 \neq \varkappa+a_{11},
\end{eqnarray}
otherwise equations (\ref{varphi})-(\ref{int}) would make no sense. 

A possible (and simple) choice for $a_{01}$ is given by: 
\begin{eqnarray}\label{ta10a01_2}
\tilde a_{01}(q) =\frac{1}{q-b},
\end{eqnarray}
implying
\begin{eqnarray}\label{ta10a01_3}
a_{01}(\mu)=  e^{i b \mu}.
\end{eqnarray}
Then, assuming that $\tilde a_{10}(q)$ is an entire function, eq.(\ref{a10a01}) gives
\begin{eqnarray}\label{a10_mod}
\tilde a_{10}(w) = 
-\frac{(w+a_{11})(w+b)}{2\pi} + \tilde a_{10}(-b), 
\end{eqnarray}
while eq.(\ref{eta_neq}) reads
\begin{eqnarray}
\eta(\varkappa)=2\pi \frac{\tilde a_{10}(-b)  }{\varkappa+b} \neq  \varkappa +a_{11} 
\;\;\Rightarrow\;\; \Omega(\varkappa)\ne 0.
\end{eqnarray}
Hereafter we assume, without loss of generality, that  
\begin{eqnarray}\label{eta_0}
\tilde a_{10}(-b)=0\;\;\Rightarrow \;\; \eta(\varkappa)=0,\;\;\Omega(\varkappa)=\varkappa + a_{11}.
\end{eqnarray}

Now we rewrite eq.(\ref{int}) using eqs.(\ref{a10a01}-\ref{a10_mod}):
\begin{eqnarray}\label{int1}
\ba{l}
\tilde \phi \Big(-\xi,x_1-\rho^{(1)}(\xi) t,\dots,x_N-\rho^{(N)}(\xi) t \Big)
\Big[\tilde  u(\xi;x)- \\
\left(\int\limits_{-\infty}^{\infty}dq
\frac{(q-b)(q-a_{11})}{(a_{11}-\xi)(q-\xi)}\tilde u(q;x)+\frac{2\pi}{a_{11}-\xi}\right)\delta(\xi-b)\Big]=0.
\ea
\end{eqnarray}
Using again equations (\ref{ta10a01_3}), eq.(\ref{uu1_mod}) yields $u$ in terms of $w$: 
\begin{eqnarray}\label{u_mod_m}
u(\lambda;x)= \frac{e^{i b \lambda}  - e^{iw \lambda} }{w-b},   
\end{eqnarray}
so that 
\begin{eqnarray}\label{tu_F}
\tilde u(q;x)= 2\pi \frac{\delta(q-b)  -  \delta(q-w)}{w-b}.   
\end{eqnarray}
Substituting it into the  eq.(\ref{int1})
%, and using again (\ref{a10a01}-\ref{a10_mod}), 
one gets:
\begin{eqnarray}\label{int2}
\frac{\delta(w-\xi)}{w-b}\tilde \phi\Big(-\xi ,x_1-\rho^{(1)}(\xi) t,\dots,
 x_N-\rho^{(N)}(\xi) t \Big)=0,
\end{eqnarray}
which is satisfied iff
\begin{eqnarray}\label{tphi}
\tilde\phi\Big(-w , x_1-\rho^{(1)}(w) t,\dots,x_N-\rho^{(N)}(w) t \Big)=0.
\end{eqnarray}

The implicit eq.(\ref{tphi}) for $w$ suggests to take function $\tilde\phi(y,x_1,\dots,x_N)$ in the form:
\begin{eqnarray}\label{ch0}
\tilde\phi(y,x_1,\dots,x_N)= y + F(x_1,\dots,x_N).
\end{eqnarray}
Thus eq.(\ref{tphi}) yields
\begin{eqnarray}\label{ch}
w=F\Big(x_1-\rho^{(1)}(w) t,\dots,x_N-\rho^{(N)}(w) t\Big),
\end{eqnarray}
which is the well-known non-differential equation defining implicitly the solution of the 
Cauchy problem in $\RR^N$
\begin{eqnarray}\label{C}
w_t + \sum_{j=1}^N w_{x_j} \rho^{(j)}(w) =0,\;\;\; w|_{t=0}=F(x_1,\dots,x_N)
\end{eqnarray}
for the scalar version of equation (\ref{3nl_1}).

Then the direct problem, the mapping from $w|_{t=0}$ to the dressing function $\psi$ 
(or $\tilde\phi$, via (\ref{varphi}) and (\ref{psi_tilde})), is simply given by
\beq
w|_{t=0}=F(x_1,\dots,x_N)\;\;\Rightarrow\;\;\tilde\phi(y,x_1,\dots,x_N)=y+F(x_1,\dots,x_N).
\eeq
From the inverse problem point of view, given $\tilde \phi$ from the initial data through 
eqs.(\ref{tphi},\ref{ch0}),  the spectral function $\tilde u$ is reconstructed solving 
equation (\ref{int1}), which is equivalent to 
\begin{eqnarray}\label{int3}
&&
 \tilde  u(\xi;x)-\int\limits_{-\infty}^{\infty}\frac{dqd\nu}{2\pi}e^{i\nu(b-\xi)}
\frac{(q-b)(q-a_{11})}{(a_{11}-\xi)(q-\xi)}\tilde  u(q;x)=\frac{2\pi}{a_{11}-\xi}\delta(\xi-b)+
 \alpha(x)\delta(w-\xi ),
\end{eqnarray}
where $\alpha(x)$ is found requiring that  
$\tilde u$ be compatible with the expression (\ref{tu_F}). This request fixes the value $\alpha=-2\pi/(w-b)$.

%%%%%%%%%%%%%%%%%%%%%%%%%%
\paragraph{Matrix nonlinear PDEs}
\label{Matrix}

In the matrix case, we choose the operators ${\cal{A}}*$ and $*A$ to be scalar; 
i.e., $a_{10}(\lambda)$ and $a_{01}(\lambda)$ are scalar functions and 
$a_{11}$ is a scalar parameter. Now we consider the general case $u_1\neq I$.
Then eq.(\ref{uu1_mod}) yields 
\begin{eqnarray}\label{uu1_mod_m}
u(\lambda;x)=\frac{u_1(x)}{2\pi i}\oint\limits_{\Gamma_{01}}d q \;
 \tilde a_{01}(q) \Big(w(x)-qI\Big)^{-1} 
e^{i q \lambda} .
\end{eqnarray}
Assuming the invertibility of $u_1$, eq.(\ref{condition_a})  yields the following 
matrix generalization of (\ref{a10a01}): 
\begin{eqnarray}\label{condition_a_m}
\frac{1}{2\pi i}\int\limits_{-\infty}^\infty d\mu \oint\limits_{\Gamma_{01}}d q\; 
a_{10}(\mu)\tilde a_{01}(q) e^{i q\mu}
 \Big(w(x)- qI\Big)^{-1}   = w(x) - a_{11}I. 
\end{eqnarray}
The expressions for $\eta$, $\tilde a_{01}(q)$ and $a_{01}(\mu)$ remain the same, 
see eqs.(\ref{eta_neq}-\ref{eta_0}).

We also assume that $w$ is diagonalizable with eigenvalues $\{e^{(j)},\dots,e^{(Q)} \}$, and that 
$V(x)$ is the matrix of right eigenvectors, so that:
\begin{eqnarray}
w=V {\mbox{diag}}(e^{(1)},\dots, e^{(Q)}) V^{-1}.
\end{eqnarray} 
Then, applying $V^{-1}$ and $V$ respectively from the left and from the right to equation 
(\ref{condition_a_m}), 
we obtain the scalar equation (\ref{a10a01}), in which $w$ is replaced by its eigenvalues. Consequently,  
also the expression of $\tilde a_{10}$ is given by (\ref{a10_mod}), and (\ref{eta_0}) holds too.

The matrix generalizations of equations (\ref{int1}), (\ref{u_mod_m}) and (\ref{tu_F}) read, respectively,
\begin{eqnarray}\label{int4b}
\ba{l}
\tilde \phi \Big(-\xi,x_1-\rho^{(1)}(\xi) t,\dots,x_N-\rho^{(N)}(\xi) t \Big)
\Big[\tilde  u(\xi;x)- \\
\left(\int\limits_{-\infty}^{\infty}dq
\frac{(q-b)(q-a_{11})}{(a_{11}-\xi)(q-\xi)}\tilde u(q;x)+
\frac{2\pi}{a_{11}-\xi}u_1(x)\right)\delta(\xi-b)\Big]=0,
\ea
\end{eqnarray}
\begin{eqnarray}\label{u_mod_mm}
u(\mu;x)=u_1(x)V(x) \Big(e^{i b \mu I} - e^{i \mu E}\Big) (E-b I)^{-1}V^{-1}(x)=
u_1(x)(e^{i b \mu I} - e^{i \mu w})(w-b I)^{-1},  
\end{eqnarray}
and
\begin{eqnarray}\label{u_mod_mmm}
\ba{l}
\tilde u(q;x)=2\pi u_1(x)V(x) \Big( I \delta(q-b) - \delta(qI-E) \Big) (E-b I)^{-1} V^{-1}(x)= \\
2\pi u_1(x)\Big( I \delta(q-b) - \delta(qI-w) \Big) (w-b I)^{-1}.  
\ea
\end{eqnarray}
Substituting (\ref{u_mod_mmm}) into the integral equation (\ref{int4b}) and multiplying 
the result by the matrix $(w-bI)$ from the right, one obtains the matrix distribution equation:
\begin{eqnarray}\label{basic_w}
\ba{l}
\tilde \phi \Big(-\xi,x_1-\rho^{(1)}(\xi) t,\dots,x_N-\rho^{(N)}(\xi) t \Big)u_1(x)\;\delta(\xi I-w)=0.
\ea
\end{eqnarray}
Since $\delta(\xi I-w)=V\delta(\xi I-E)V^{-1}$, equation (\ref{basic_w}) 
is equivalent to the distribution equation 
\begin{eqnarray}\label{basic2}
\ba{l}
\sum_{\gamma_1=1}^Q  \sum_{\gamma_2=1}^Q \sum_{\gamma_3=1}^Q
\tilde \phi_{\alpha\gamma_1}\Big(-\xi,x_1-\rho^{(1)}(\xi) t,\dots, x_N-\rho^{(N)}(\xi) t \Big)\times \\
\;\; \\
(u_1(x))_{\gamma_1\gamma_2} V_{\gamma_2\gamma_3}(x)\delta(\xi-e^{(\gamma_3)})V^{-1}_{\gamma_3\beta}(x)
=0,
\ea
\end{eqnarray}
which implies 
 \begin{eqnarray}\label{matr_w}
\ba{l}
\sum_{\gamma_1=1}^Q  \sum_{\gamma_2=1}^Q \sum_{\gamma_3=1}^Q
\tilde \phi_{\alpha\gamma_1}(-e^{(\gamma_3)} ,x_1-\rho^{(1)}(e^{(\gamma_3)}) t,\dots, x_N-
\rho^{(N)}(e^{(\gamma_3)}) t ) 
\times \\ 
(u_1(x))_{\gamma_1\gamma_2} V_{\gamma_2\gamma_3}(x)V^{-1}_{\gamma_3\beta}(x)
=0,\;\;\;\alpha,\beta=1,\dots,Q.
\ea
\end{eqnarray}
Choosing, in analogy with the scalar case, the following form for the matrix $\tilde\phi$:
\begin{eqnarray}\label{int4}
\tilde \phi_{ij} \Big(y,x_1,\dots,x_N \Big)=y\delta_{ij}+F_{ij}\Big(x_1,\dots,x_N \Big) ,
\end{eqnarray}
and eliminating eigenvalues and eigenvectors, equation (\ref{matr_w}) becomes
 \begin{eqnarray}\label{matr_ww}
(u_1w)_{\alpha\beta}= \sum_{\gamma_1=1}^Q  \sum_{\gamma_2=1}^Q
(u_1(x))_{\gamma_1\gamma_2}\Big(F_{\alpha\gamma_1}(x_1 I-\rho^{(1)}(w) t,\dots,
 x_N I-\rho^{(N)}(w) t )\Big)_{\gamma_2\beta}.
\end{eqnarray}
Applying $u^{-1}_1$ from the left we finally obtain the non-differential equation (\ref{sol1_w}), involving the 
solution $w(x)$ of (\ref{3nl_1}) and the auxiliary matrix function $u_1(x)$.   

%%%%%%%%%%%%%%%
\subsubsection{The dressing function $G(\lambda;x)$ and the construction of $u_1(x)$}
\label{FunctionG}\label{RHS}
%\paragraph{ Construction of $u_1$}
If $T\neq 0$, the solution of (\ref{3nl_1}) is computed involving also functions $u_1$ and $G$, 
which are related by equation (\ref{SPG}). 
Substituting $u$ given by (\ref{u_mod_mm}) into (\ref{SPG})  and using the following Fourier representation 
for $g(\mu;x)$:
 \begin{eqnarray}
 g(\mu;x)=\int\limits_{-\infty}^\infty \frac{d\varkappa}{2\pi} \; \hat g(\varkappa;x) e^{i \varkappa \mu},
 \end{eqnarray}
one obtains 
\begin{eqnarray}\label{gu1}
&&
\sum_{\gamma_1=1}^Q  \sum_{\gamma_2=1}^Q \Big[
\Big(\hat g_{\alpha\gamma_1}(-b;x) I - 
\hat g_{\alpha\gamma_1} (-w;x)\Big)
(w-bI)^{-1}
\Big]_{\gamma_2\beta} (u_1(x))_{\gamma_1\gamma_2} +\\\nonumber
&& 
(g_1(x)
 u_1(x))_{\alpha\beta}
= 
\delta_{\alpha\beta},\;\;\alpha,\beta=1,\dots,Q.
\end{eqnarray}
This is a determined system of linear equations for the elements of $u_1$,
where functions $g$ and $g_1$ are solutions of eq.(\ref{G}), which 
reads, in terms of (\ref{A_def},\ref{G_def}) and (\ref{A_def_j}):
\begin{eqnarray}\label{Gg_ex21}
&&
g_{t}(\lambda;x) + 
\sum_{j=1}^N g_{x_j}(\nu;x)*a^{(j)}(\nu,\lambda)+ \sum_{j=1}^N
{g_1}_{x_j}(x)a^{(j)}_{10}(\lambda)  = \\\nonumber
&&
- T g(\nu;x)*a^{(0)}(\nu,\lambda) - 
T g_1(x) a^{(0)}_{10}(\lambda),\;\;\;\;\lambda\in {\cal{D}}, \\\label{Gg_ex22}
&&
{g_1}_{t}(x) + \sum_{j=1}^N {g_1}_{x_j}(x) a^{(j)}_{11} + 
\sum_{j=1}^Ng_{x_j}(\mu;x)*a^{(j)}_{01}(\mu) =\\\nonumber
&& - T g_1(x)a^{(0)}_{11} - 
T g(\mu;x)*a^{(0)}_{01}(\mu),\;\;\;\;\lambda=l_1.
\end{eqnarray}
The representations of functions $A^{(j)},~j=0,1,2,..$ in terms of $A$ 
can be obtained recursively through the formulae    
\begin{eqnarray}\label{recursion}
A^{(n)}(\lambda,\nu)* A(\nu,\mu)=
\left\{
\begin{array}{ll}
a^{(n)}(\lambda,\nu)*a(\nu,\mu)+ a^{(n)}_{01}(\lambda) a_{10}(\mu) ,& \lambda,\mu\in {\cal{D}} , \cr
a^{(n)}_{01}(\lambda)a_{11}+ a^{(n)}(\lambda,\nu) * a_{01}(\nu) ,& \lambda\in {\cal{D}},\;\;\mu=l_1 , \cr
a^{(n)}_{10}(\nu)*a(\nu,\mu)+ a^{(n)}_{11} a_{10}(\mu) ,& \lambda=l_1,\;\;\mu\in {\cal{D}} , \cr
a^{(n)}_{10}(\nu)*a_{01}(\nu)+ a^{(n)}_{11} a_{11} ,& \lambda=\mu=l_1 .
\end{array}\right. 
\end{eqnarray}
Due to the choice (\ref{a_ex2}) for $a$, and to the representations (\ref{a1001},\ref{ta10a01_2},\ref{ta10a01_3}) 
of $a_{10}$ and $a_{01}$, it is possible to  
look for a solution of the system  (\ref{Gg_ex21}, \ref{Gg_ex22}) satisfying the following properties:
\begin{eqnarray}\label{g1g}
\ba{l}
g_1=0,\;\; g(\lambda;x)*a^{(j)}_{01}(\lambda) =0, \\
g(\mu;x)*a^{(j)}(\mu,\lambda) = g(\mu;x)*\rho^{(j)}(a(\mu,\lambda)) 
\equiv  \rho^{(j)}(i\partial_\lambda) g(\lambda;x) ,\;\;j=1,2,\dots
\ea
\end{eqnarray}
which reduce such system to the single equation 
\begin{eqnarray}
\label{Gg_ex23}
g_t(\lambda;x) +
\sum_{j=1}^N \rho^{(j)}(i\partial_\lambda)\; g_{x_j}(\lambda;x) +T \rho^{(0)}(i\partial_\lambda)\; g(\lambda;x)=0.
\end{eqnarray}
Indeed, as it is possible to verify using equation $a_{01}(\mu)=\exp(ib\mu)$ and formulae (\ref{recursion}), 
the restrictions 
(\ref{g1g}) are simply satisfied by imposing the single condition $\hat g(-b;x)=0$. 

Then the matrix function $g$ satisfying (\ref{g1g},\ref{Gg_ex23}) can be represented in the following 
Fourier form:
\begin{eqnarray}\label{gexp}
\ba{l}
g(\lambda;x)=\int\limits_{-\infty}^\infty \frac{d \varkappa}{2\pi}\int\limits_{\RR^N} 
dk\int\limits_{\Gamma_{\omega}} d\omega\;\tilde g(\varkappa,k,\omega) 
e^{ i \varkappa \lambda + i \sum_{j=1}^N k_jx_j  -i \omega t}, \\
\tilde g_{\alpha\beta}(\varkappa,k,\omega)=(\varkappa+b)h_{\alpha\beta}(k)
\delta\Big(\omega-\sum_{j=1}^N k_j \rho^{(j)}(-\varkappa)+i\rho^{(0)}(-\varkappa)T_{\alpha}\Big)
\ea
\end{eqnarray}
where $\Gamma_{\omega}$ is any contour passing through the support of the Dirac function,   
and $h_{\alpha\beta},\;\alpha,\beta=1,\dots,Q$ are arbitrary scalar functions of their arguments. 

Then replacing 
\begin{eqnarray}
 \hat g(\varkappa;x)=\int\limits_{\RR^N}
dk\int\limits_{\Gamma_{\omega}} 
d\omega\; \tilde g(\varkappa,k,\omega) 
e^{ i \sum_{j=1}^N k_j x_j -i \omega t}.
\end{eqnarray}
in equation (\ref{gu1}), one obtains:
\beq\label{pre_sol1}
\sum_{\gamma_1,\gamma_2=1}^Q\Big[
H_{\alpha\gamma_1}\big(x_1I-\rho^{(1)}(w)t,\dots,x_NI-\rho^{(N)}(w)t\big)e^{-\rho^{(0)}(w)T_{\alpha}t}\Big]_{\gamma_2\beta}
(u_1(x))_{\gamma_1\gamma_2}= \delta_{\alpha\beta} ,
\eeq
where
\beq
H_{\alpha\beta}(x_1,\dots,x_N)\equiv \int_{\RR^N}dk \; h_{\alpha\beta}(k)e^{i\sum_{l=1}^N k_l x_l}.
\eeq

If we choose $h_{\alpha\beta}(\xi,k)=(\prod_j\delta(k_j))\delta_{\alpha\beta}$, 
$\Rightarrow$ $H_{\alpha\beta}=\delta_{\alpha\beta}$; then equation (\ref{pre_sol1})  
reduces to (\ref{sol1_u1}). This simplification can be done without loss of generality   
since, if it were not made, it would correspond to the following redefinition of the arbitrary matrix function $F$ 
appearing in (\ref{sol1_w}):  
\beq
F\to HFH^{-1}. \;\;\;\;\;\;\Box
\eeq

%%%%%%%%%%%%%%%%%%%%%%%%%%%

%%%%%%%%%%%%%%%%%%%%
\section{Second order quasilinear PDEs}
\label{Second order PDEs}
%%%%%%%%%%%%%%%%%%%%%%%%%%%%
 \subsection{Derivation of second order PDEs}
\label{Higher_order}\label{Second_order}
To increase the order of the nonlinear PDEs, we 
introduce the $x$-dependence through the next equations
\begin{eqnarray}\label{x2}
&&
\Psi_{t_m}(\lambda,\mu;x)+
\sum_{j=1}^{N} {\cal{A}}^{(mj)}(\lambda,\nu)*\Psi_{x_j}(\nu,\mu;x) = 0,\\\label{G2}
&&
G_{t_m}(\lambda,q;x)+
\sum_{j=1}^{N} G_{x_j}(\nu,q;x)*A^{(mj)}(\nu,\lambda)
 = -q^m S^{(m)} G(\nu,q;x)*A^{(m0)}(\nu,\lambda) ,  
\end{eqnarray}
supplemented by the generalized commutation relation:
\begin{eqnarray}\label{commj_2}
{\cal{A}}^{(mj)}(\lambda,\nu)* \Psi(\nu,\mu;x) = \Psi(\lambda,\nu;x)*
 A^{(mj)}(\nu,\mu),
\end{eqnarray}
where $q$  is a new spectral parameter, $S^{(m)}$ are constant diagonal matrices, 
\begin{eqnarray}\label{commj_22}
{\cal A}^{(mj)}*=\rho^{(mj)}({\cal A})*,\;\;\;*A^{(mj)}=*\rho^{(mj)}(A),
\end{eqnarray}
and $\rho^{(mj)}:\;\RR\to\RR$ are arbitrary analytic functions. 

Together  with the field $w$ introduced in the previous section, we introduce also the fields 
\begin{eqnarray}\label{def_v}
v^{(n)}(x)=(G(\mu,q;x)q^n) * U(\mu;x),\;\;n=1,2,\dots,\;\;v(x)\equiv v^{(1)}(x) ,
\end{eqnarray}
where now "$*$" means integration also over $q$.

Applying the operators $\Big(\partial_{t_m}+\sum_{j=1}^N {\cal{A}}^{(mj)} \partial_{x_j} * \Big)$ to 
(\ref{U}), one gets
\begin{eqnarray}
\Psi(\lambda,\mu)*\Big(U_{t_m}(\mu;x) + 
\sum_{j=1}^N A^{(mj)}(\mu,\nu)*U_{x_j}(\nu,\mu;x)\Big) = 0.
\end{eqnarray}
Assuming, as before, that the solutions of (\ref{U}) belong to the 
 one-dimensional matrix subspace generated by $U$, the spectral equations are similar to 
(\ref{3sp_2},\ref{3sp_1}): 
\begin{eqnarray}\label{2_3sp_2}
&&
A(\lambda,\nu)*U(\nu;x) = U(\lambda;x) \tilde F(x),\\
\label{2_3sp_1}
&&
U_{t_m}(\lambda;x) + 
\sum_{j=1}^N A^{(mj)}(\lambda,\nu)*U_{x_j}(\nu,\mu;x)=
U(\lambda;x) F^{(m)}(x) .
\end{eqnarray}
Applying $G*$ to (\ref{2_3sp_2}) and (\ref{2_3sp_1}) and using the condition (\ref{condition}) 
and equations (\ref{G2}), (\ref{AiU}), we obtain the expression of $\tilde F(x)$ and  
$F^{(m)}(x)$ in terms of the matrix fields $w$ and $v^{(n)}$:
\begin{eqnarray} 
\tilde F(x)= w(x),\;\;\;
F^{(m)}(x)=\sum_{j=1}^N \left(\rho^{(mj)}(w)\right)_{x_j}+ S^{(m)}
v^{(m )}(x)\rho^{(m0)}(w).
\end{eqnarray}
Therefore equations (\ref{2_3sp_2}) and (\ref{2_3sp_1}) become the following 
overdetermined system for the spectral function $U(\lambda;x)$:  
\begin{eqnarray}\label{2_3sp_13}
\ba{l}
A(\lambda,\mu)U(\mu;x)=U(\lambda;x)w(x), \\
U_{t_m}(\lambda;x) + 
\sum\limits_{j=1}^N U_{x_j}(\lambda;x) \rho^{(mj)}(w)=U(\lambda;x)  S^{(m)}v^{(m)}(x)\rho^{(m0)}(w) .
\ea
\end{eqnarray}
In order to construct the nonlinear PDE for $w$, we apply $G*A*$ to equations (\ref{2_3sp_13}b) 
and use equations  (\ref{AiU}), (\ref{x2}) and (\ref{def_v}), obtaining
\begin{eqnarray}\label{2_3nl_13}
w_{t_m} + 
\sum_{j=1}^N w_{x_j} \rho^{(mj)}(w)=[w, S^{(m)}v^{(m)}\rho^{(m0)}(w)].
\end{eqnarray}

To write the equations for $v^{(k)}$, we apply instead $(Gq^k)*$ to eq.(\ref{2_3sp_13}b), obtaining:
\begin{eqnarray}
   \label{2_nl2}
v^{(k)}_{t_m} + 
\sum_{j=1}^N v^{(k)}_{x_j} \rho^{(mj)}(w)+S^{(m)} v^{(k+m)} \rho^{(m0)}(w)=
v^{(k)} S^{(m)} v^{(m)}\rho^{(m0)}(w).
\end{eqnarray}
To construct a complete system of nonlinear PDEs, consider the three equations:
 \begin{eqnarray}
   \label{2_nlf}
v^{(1)}_{t_1} + 
\sum_{j=1}^N v^{(1)}_{x_j} \rho^{(1j)}(w)+S^{(1)} v^{(2)}\rho^{(10)}(w) =
v^{(1)} S^{(1)} v^{(1)}\rho^{(10)}(w),\\\nonumber 
v^{(2)}_{t_1} + 
\sum_{j=1}^N v^{(2)}_{x_j} \rho^{(1j)}(w)+S^{(1)} v^{(3)}\rho^{(10)}(w) =
v^{(2)} S^{(1)} v^{(1)}\rho^{(10)}(w),\\\nonumber
v^{(1)}_{t_2} + 
\sum_{j=1}^N v^{(1)}_{x_j} \rho^{(2j)}(w)+S^{(2)} v^{(3)}\rho^{(20)}(w) =
v^{(1)} S^{(2)} v^{(2)}\rho^{(20)}(w),
\end{eqnarray}
corresponding to $(m,k)=(1,1), (1,2), (2,1)$. 

We may choose, without loss of generality, $S^{(1)}=1$, $S^{(2)}=S$. 
Then, expressing $v^{(2)}$ and $v^{(3)}$ in terms of $v=v^{(1)}$:
\beq
\ba{l}
v^{(2)}=v^2-{\cal{L}}_1(v)(\rho^{(10)}(w))^{-1}, \\
v^{(3)}=v^{(2)}v-{\cal{L}}_1 (v^{(2)})(\rho^{(10)}(w))^{-1},
\ea
\eeq
one obtains a single matrix equation of the second order for $v$ and $w$:
\beq\label{v_equ}
\ba{l}
{\cal{L}}_2(v) \rho^{(10)}(w) +  S{\cal{L}}_1 \Big( {\cal{L}}_1 (v)(\rho^{(10)}(w))^{-1} \Big) \rho^{(20)}(w)   
= S{\cal{L}}_1(v^2) \rho^{(20)}(w)+  
\left[ S {\cal{L}}_1(v) \rho^{(20)}(w) , v\right] + \\
S{\cal{L}}_1(v) \left[(\rho^{(10)}(w))^{-1}v,\rho^{(20)}(w)\rho^{(10)}(w)\right]+
\left[v,Sv^2\right]\rho^{(20)}(w)\rho^{(10)}(w) ,
\ea
\eeq
where the differential operators ${\cal{L}}_m,\;m=1,2$ are defined by
\begin{eqnarray}
{\cal{L}}_m(f(x)) = f_{t_m}(x) + 
\sum_{i=1}^N  f_{x_i}(x) \rho^{(mi)}(w) .
\end{eqnarray}

A complete system of second order matrix PDEs in arbitrary dimensions for the fields $w$ and $v$ is  
obtained coupling equation (\ref{v_equ}) with equation 
(\ref{2_3nl_13}) for $m=2$, obtaining:
\beq\label{B}
\ba{l}
{\cal{L}}_2(w)=[w,S(v^2-{\cal{L}}_1(v))v(\rho^{(10)}(w))^{-1}], \\
{\cal{L}}_2(v) \rho^{(10)}(w) +  S{\cal{L}}_1 \Big( {\cal{L}}_1 (v)(\rho^{(10)}(w))^{-1} \Big) \rho^{(20)}(w)   
= S{\cal{L}}_1(v^2) \rho^{(20)}(w)+  
\left[ S {\cal{L}}_1(v) \rho^{(20)}(w) , v\right] + \\
S{\cal{L}}_1(v) \left[(\rho^{(10)}(w))^{-1}v,\rho^{(20)}(w)\rho^{(10)}(w)\right]+
\left[v,Sv^2\right]\rho^{(20)}(w)\rho^{(10)}(w) .
\ea
\eeq
But the independent equation (\ref{2_3nl_13}) for $m=1$:
\beq\label{A}
{\cal{L}}_1(w)=[w,v\rho^{(10)}(w)] 
\eeq
is also satisfied by the fields $w$ and $v$ and must be viewed as an integrable 
constraint for the evolutionary (with respect to the time $t_2$) system (\ref{B}). 

We remark that the evolutionary system in arbitrary dimensions (\ref{B}) is completely integrable 
only under the nonlinear constraint (\ref{A}). 
 
Here we consider two explicit reductions.

1. Let
\begin{eqnarray}\label{delta_red}
N=2,\;\;\rho^{(11)}=\rho^{(22)}= w,\;\;\rho^{(10)}=\rho^{(20)}=I,\;\;\;\rho^{(12)}=\rho^{(21)}= 0,
\end{eqnarray}
then
\begin{eqnarray}\label{delta_red2}
{\cal{L}}_1(f)= f_{t_1} + f_{x_1} w,\;\;{\cal{L}}_2(f)= f_{t_2} + f_{x_2} w,
\end{eqnarray}
and one obtains the nonlinear system:
\begin{eqnarray}\label{2_nlf_red1}
&&
w_{t_2} + 
  w_{x_2}w=[w,S (v^2-v_{t_1} - v_{x_1} w)] , \\\nonumber
&&
 v_{t_2} +v_{x_2} w  +  
S(v_{t_1}+
v_{x_1} w)_{t_1}+S(v_{t_1}+
v_{x_1} w)_{x_1} w    =\\\nonumber
&&
S\Big((v^2)_{t_1} +(v^2)_{x_1} w\Big)  +
[S(v_{t_1}+v_{x_1} w)   ,  v] +
[v,S v^2] ,
\end{eqnarray}
subjected to the constraint
\beq\label{2_nlf_red12}
w_{t_1} + w_{x_1}w=[w, v].
\eeq
Using this constraint, eq.(\ref{2_nlf_red1}b) can be rewritten in the more convenient form
\begin{eqnarray}
v_{t_2}+ Sv_{t_1t_1} + v_{x_2} w + Sv_{x_1x_1} w^2 + 2Sv_{x_1t_1} w -[S,v] (v_{t_1}+v_{x_1} w)  -\\\nonumber
2 S(v_{t_1} v + v_{x_1} v  w) +[Sv^2,v] =0.
\end{eqnarray}

2. Let
\begin{eqnarray}
N=2,\;\;\rho^{(11)}=\rho^{(22)}=\rho^{(10)}=\rho^{(20)}=I,\;\;\;\rho^{(12)}=\rho^{(21)}=
0,
\end{eqnarray}
then 
\begin{eqnarray}
{\cal{L}}_1(f)= f_{t_1} + f_{x_1} =f_{\tau_1},\;\;
{\cal{L}}_2(f)= f_{t_2} + f_{x_2} =f_{\tau_2},
\end{eqnarray}
and the nonlinear system  reads
\begin{eqnarray}\label{Burgers}
&&
w_{\tau_2} =[w,S (v^2-v_{\tau_1}  )] , \\\nonumber
&&
 v_{\tau_2} + S v_{\tau_1\tau_1}  =
 S(v^2)_{\tau_1} +[ S v_{\tau_1}  ,  v] +
[v, S v^2],
\end{eqnarray}
with the constraint
\beq
w_{\tau_1} =[w, v].
\eeq
We remark that, in this reduction, equation (\ref{Burgers}b), which involves only the field $v$, 
is the matrix B\"urgers equation \cite{BLR}.

Using equations (\ref{U_def}), (\ref{G_def}) and (\ref{def_v}), the fields $v^{(j)}$ have the 
following spectral representation:
\begin{eqnarray}\label{repr_v}
v^{(j)}(x)\equiv (G(\mu,q;x) q^j)*U(\mu;x) &=& (g(\mu,q;x)q^j
 u(\mu;x)\big)+
\int g_1(q,x) q^j dq u_1.
\end{eqnarray}
%while $w$ is defined by the eq.(\ref{w_dr})

The dressing functions $\psi$ and $G$ are responsible for the dimensionality of the space 
of solutions for $w$ and $v$ respectively.

As we did in Sec.(\ref{Reductions}), we may separate diagonal and off-diagonal parts of the 
equation (\ref{2_3nl_13}), obtaining
\begin{eqnarray}\label{2_ee}
&&
e^{(k)}_{t_m} + \sum_{j=1}^N e^{(k)}_{x_j} \rho^{(mj)}(e^{(k)}) =0,\;\;k=1,\dots,Q,\\
\label{2_VV}
&&
V_{t_m}+\sum_{j=1}^N V_{x_j} \rho^{(mj)}(E) +S^{(m)}v^{(m)} V =V D,\;\;\;D \mbox{ diagonal.}
\end{eqnarray}
Equation (\ref{2_ee}) coincides with equation (\ref{ee}); i.e., the eigenvalues of $w$ evolve separately 
according to the scalar version of equation (\ref{2_3nl_13}). Instead, eq. (\ref{2_VV}) cannot be written in the  
same form of eq. (\ref{VV}), since $v^{(m)}$ is not diagonal matrix. This is a principal difference between 
eq. (\ref{2_3nl_13}) and eq. (\ref{3nl_1}). 

%%%%%%%%%%%%%%%
\subsection{The solution of the system (\ref{B}), (\ref{A})}
\label{FunctionG2}\label{RHS2}
In this section we construct, using the dressing method introduced in this paper, 
the solution of the matrix equations (\ref{A}) and (\ref{B}), which 
turns out to be characterized by a nonlinear system of non-differential equations for the 
components of the unknown matrices $w$ and $v$ in the following way.

\vskip 5pt
\noindent
{\bf Proposition 2}. Let $F_{ij}:\RR^N\to\RR,\;i,j=1,\dots,Q$ and $H_{ij}:\RR^{N+1}\to\RR,\;i,j=1,\dots,Q$ 
be $2Q^2$ arbitrary scalar functions, 
representable by positive power series, so that $F_{ij}(M_1,\dots,M_N)$ and $H_{ij}(q,M_1,\dots,M_N)$ are well defined matrix 
functions, where $M_1,\dots,M_N$ are arbitrary $Q\times Q$ matrices. Then the solutions of the 
matrix system (\ref{B}), subjected to the matrix constraint(\ref{A}), are characterized by the following system of 
$3Q^2$ non-differential equations: 
\beq\label{Sol1_w}
\ba{l}
w_{\alpha\beta} = \sum_{\gamma_1,\gamma_2,\delta=1}^Q (u_1^{-1}(x))_{\alpha\delta}
 \Big(F_{\delta\gamma_1}(x_1 I-\sum_{m=1}^2\rho^{(m1)}(w) t_m,\dots,x_N I-
\sum_{m=1}^2 \rho^{(mN)}(w) t_m)\Big)_{\gamma_2\beta} \times \\
(u_1(x))_{\gamma_1\gamma_2}=0,\;\;\;\;\;\;\;\; \alpha,\beta=1,\dots,Q ,
\ea
\eeq
\beq\label{Sol1_u1}
\ba{l}
\int\limits_{-\infty}^\infty dq\;\sum_{\gamma_1=1}^Q  \sum_{\gamma_2=1}^Q 
\Big[H_{\alpha\gamma_1}(q,x_1I -\sum_{m=1}^2 \rho^{(m1)}(w) t_m,\dots,  x_N I -\sum_{m=1}^2 \rho^{(mN)}(w) t_m)
\times \\
e^{-\sum_{m=1}^2  S^{(m)}_\alpha q^m \rho^{(m0)}(w)t_m}\Big]_{\gamma_2\beta}
 (u_1(x))_{\gamma_1\gamma_2}
= \delta_{\alpha\beta},\;\;\;\;\;\;\;\;
\alpha,\beta=1,\dots,Q,
\ea
\eeq
\beq\label{Sol1_v}
\ba{l}
v_{\alpha\beta}(x)=\int\limits_{-\infty}^\infty dq\;q\sum_{\gamma_1,\gamma_2=1}^Q
\Big(H_{\alpha\gamma_1}\left(q,x_1I -\sum_{m=1}^2 \rho^{(m1)}(w) t_m,\dots,x_N I -\sum_{m=1}^2 
\rho^{(mN)}(w) t_m\right)\times \\
e^{-\sum_{m=1}^2S^{(m)}_\alpha q^m \rho^{(m0)}(w) t_m} \Big)_{\gamma_2 \beta}
 (u_1(x))_{\gamma_1\gamma_2},
\;\;\;\;\;\;\;\;\alpha,\beta=1,\dots,Q ,
\ea
\eeq
for the components of the unknown matrices $w(x)$ and $v(x)$, and of the auxiliary matrix function $u_1(x)$. 

\vskip 5pt
\noindent
{\bf Proof of Proposition 2.} To prove this Proposition, we proceed as in Secs.\ref{First_order},\ref{Solutions}. 
The solutions $w$ and $v^{(n)}$ are constructed using the algorithm presented in Sec.\ref{Solutions}, 
but the functions $\Psi$ and $G$ will be now defined by eqs. (\ref{x2},\ref{G2}). 
Eq.(\ref{x2}) yields
\begin{eqnarray}\label{ex2:t21}
&&\psi_{t_m}(\lambda,\mu;x) +\sum_{j=1}^N \rho^{(mj)}(i\partial_\lambda) 
\psi_{x_j }(\lambda,\mu;x)=0 ,\;\;
\lambda \in {\cal{D}},\\\nonumber
\label{ex2:t22}
&&{\psi_{01}}_{t_m}(\lambda;x)  + \sum_{j=1}^N \rho^{(mj)}(i\partial_\lambda)
{\psi_{01}}_{x_j}(\lambda;x)=0,\;\; \lambda =l_1,
\end{eqnarray}
where $m=1,2$.
The analysis of the system (\ref{ex2:t21})
coincides with the analysis carried out in Sec.\ref{Psix_neq_0}.
but $\psi$ and $\psi_{01}$ exhibit the following Fourier representations:
\begin{eqnarray}\label{psi_F_2}
&&\psi(\lambda,\mu;x)=\int\limits_{-\infty}^\infty 
d\varkappa\int\limits_{-\infty}^\infty d q \int\limits_{\RR^N}dk \;
\tilde \psi(\varkappa,q,k) 
e^{i \varkappa \lambda + i q \mu + i \sum_{j=1}^N k_j \Big(x_j -
 \sum_{m=1}^2 \rho^{(mj)}(-\varkappa) t_m\Big)},\\\label{psi01_F_2}
&&\psi_{01}\lambda;x)=
\int\limits_{-\infty}^\infty  d\varkappa
\int\limits_{\RR^N}dk \;
\tilde \psi_{01}(\varkappa,k) 
e^{i \varkappa \lambda  + i \sum_{j=1}^N k_j \Big(x_j -
\sum_{m=1}^2 \rho^{(mj)}( - \varkappa) t_m \Big)} ,
\end{eqnarray}
and equation (\ref{sol1_w}) is replaced by (\ref{Sol1_w}).
 
The $x$-dependence of $G$ is introduced by eq.(\ref{G2}), with $a(\lambda,\mu)$ given in (\ref{a_ex2}):  
\begin{eqnarray}\label{G_mod3}
&&
g_{t_m}(\lambda,q;x) + 
\sum_{j=1}^N g_{x_j}(\nu,q;x)*a^{(mj)}(\nu,\lambda)+ 
\sum_{j=1}^N {g_1}_{x_j}(q,x)a^{(mj)}_{10}(\lambda)  = \\\nonumber
&&
- S^{(m)} q^m \Big(g(\nu,q;x)*a^{(m0)}(\nu,\lambda)+ 
 g_1(q,x)a^{(m0)}_{10}(\lambda)\Big),\;\;\;\;\lambda\in {\cal{D}}\\\nonumber
&&
{g_1}_{t_m}(q,x) + \sum_{j=1}^N {g_1}_{x_j}(q,x) a^{(mj)}_{11} + 
\sum_{j=1}^Ng_{x_j}(\mu,q;x)*a^{(mj)}_{01}(\mu) =\\\nonumber
&& - S^{(m)}q^m \Big(g_1(q,x) a^{(m0)}_{11} + 
 g(\mu,q;x)*a^{(m0)}_{01}(\mu)\Big),\;\;\;\;\lambda=l_1,
\end{eqnarray}
where $m=1,2$. We impose again the conditions 
\begin{eqnarray}\label{g1g2}
&&
g_1(q;x)=0,\;\; g(\mu,q;x)*a^{(mj)}_{01}(\mu) =0,\;\;\\\nonumber
&&
g(\mu,q;x)*a^{(mj)}(\mu,\lambda) = g(\mu,q;x)*\rho^{(mj)}(a(\mu,\lambda)) 
= \rho^{(mj)}(i\partial_\lambda) g(\lambda,q;x) ,\;\;j=1,2,\dots
\end{eqnarray}
on the solution of the system (\ref{G_mod3}), reducing it to the single equation
 \begin{eqnarray}\label{G_mod4}
&&
g_{t_m}(\lambda,q;x) +
\sum_{j=1}^N \rho^{(mj)}(i\partial_\lambda)\;g_{x_j}(\lambda,q;x)+S^{(m)} q^m 
\rho^{(m0)}(i\partial_\lambda)\;g(\lambda,q;x)=0.
 \end{eqnarray}
 Functions $g$ satisfying (\ref{g1g2},\ref{G_mod4}) 
 can be written in the following Fourier form:
 \begin{eqnarray}\label{g_Fourier}
&&g(\lambda,q;x)=\int\limits_{\RR^N}
dk\int\limits_{-\infty}^\infty 
d\omega
\int\limits_{-\infty}^\infty d \varkappa\; \tilde g(\varkappa,q,k,\omega) 
e^{ i \varkappa \lambda + i \sum_{l=1}^N k_{l} x_l -
 i \sum_{m=1}^2\omega^{(m)} t_m}
\end{eqnarray}
where
\begin{eqnarray}\label{sol_tg12}
\tilde g_{\alpha\beta}(\varkappa,q,k,\omega)=
 (\varkappa+b) h_{\alpha\beta}(q,k)\prod_{m=1}^2 \delta \Big(\omega^{(m)} -
 \sum_{l=1}^N k_{l} \rho^{(ml)}(-\varkappa)  +
i S^{(m)}_\alpha q^m \rho^{(m0)}(-\varkappa)\Big).
\end{eqnarray}
Now $\hat g(\varkappa,q;x)$ takes the form:
\begin{eqnarray}
&&\hat g(\varkappa,q;x)=\int\limits_{\RR^N}
dk\int\limits_{-\infty}^\infty 
d\omega
\; \tilde g(\varkappa,q,k,\omega) 
e^{ i \sum_{l=1}^N k_{l} x_l -
 i \sum_{m=1}^2\omega^{(m)} t_m},
\end{eqnarray}
and the analogue of equation (\ref{sol1_u1}) is equation (\ref{Sol1_u1}),  
where 
\beq
H(q, x_1 ,\dots, x_N)=\int\limits_{\RR^N} d k\; h(q,k) e^{i\sum_{j=1}^N k_{j}x_j} .
\eeq
At last, substituting in equation (\ref{repr_v}) the expressions (\ref{u_mod_mm}) and (\ref{g_Fourier}) 
of $u$ and $g$, we obtain (\ref{Sol1_v}). $\Box$

%%%%%%%%%%%%%%%%%%%%%%%%
\paragraph{Initial-Boundary Value Problem.}
A well-posed initial-boundary value problem for equations (\ref{B}), (\ref{A}) is the 
Cauchy problem for the system (\ref{B}), in which the initial conditions for $w$ and $v$ at $t_2=0$ 
are any pair of matrix functions satisfying the multidimensional PDE (\ref{A}). 
This initial constraint can be satisfied assigning arbitrarily $v$ at $t_2=0$ and $w$ at $t_1=t_2=0$:
\beq
w^{(00)}=w(x)|_{t_1=t_2=0}\;\;\;\;v^{(0)}=v(x)|_{t_2=0}. 
\eeq
Then, integrating equation (\ref{A}) with respect to $t_1$, one obtains $w^{(0)}=w(x)|_{t_2=0}$, $\forall t_1$. 
At last, given $v^{(0)},w^{(0)}$, the evolutionary system (\ref{B}) allows one to construct $w$ and $v$, $\forall t_2$.

The algorithm allowing one to integrate such an initial-boundary value problem consists of three steps.

1. At  $t_1=t_2=0$, the system  (\ref{Sol1_w},\ref{Sol1_u1},\ref{Sol1_v}) reads:
\begin{eqnarray}
&&
w^{(00)} = (u^{(00)})^{-1} F(x_1,\dots,x_N) u^{(00)},\\
&&
\tilde H(0,x_1,\dots,x_N) u^{(00)}=I,\\
&&
v^{(00)}=\tilde H'(0,x_1,\dots,x_N)u^{(00)} ,
\end{eqnarray} 
where $v^{(00)}=v^{(0)}(x)|_{t_1=0}$ is given, $u^{(00)}=u_1(x)|_{t_1=t_2=0}$, and 
\beq
\ba{l}
\tilde H(t_1,x_1,\dots,x_N)=\int\limits_{-\infty}^\infty dq\; H(q,x_1,\dots,x_N) e^{-qt_1} , \\
\tilde H'(0,x_1,\dots,x_N)=\tilde H_{t_1}(t_1,x_1,\dots,x_N)|_{t_1=0}.
\ea
\eeq
This system of three matrix equations must be solved for $F(x_1,\dots,x_N)$, 
$u^{(00)}$ and  $\tilde H'(0,x_1,\dots,x_N)$. Function $\tilde H(0,x_1,\dots,x_N)$ remains arbitrary.

2. The system (\ref{Sol1_w},\ref{Sol1_u1},\ref{Sol1_v}), evaluated at $t_2=0$, reads:
 \begin{eqnarray}\label{t2_matr_w22}
 w^{(0)}_{\alpha\beta} = \sum_{\gamma_1,\gamma_2,\delta=1}^Q((u^{(0)})^{-1})_{\alpha\delta} 
 \Big[F_{\delta\gamma_1}(x_1 I - \rho^{(11)}(w^{(0)}) t_1,\dots,x_N I - \rho^{(1N)}(w^{(0)}) t_1 
  )\Big]_{\gamma_2\beta} (u^{(0)})_{\gamma_1\gamma_2},\\\nonumber
  \alpha,\beta=1,\dots,Q ,
  \end{eqnarray}
\begin{eqnarray}\label{t2_gu3}
\sum_{\gamma_1=1}^Q  \sum_{\gamma_2=1}^Q 
\Big[\tilde H_{\alpha\gamma_1}(\rho^{(10)}(w^{(0)})t_1,x_1 I - \rho^{(11)}(w^{(0)}) t_1,\dots,\\\nonumber
x_N I - \rho^{(1N)}(w^{(0)}) t_1  )\Big]_{\gamma_2\beta}
 (u^{(0)})_{\gamma_1\gamma_2}
= 
\delta_{\alpha\beta},\;\;
\alpha,\beta=1,\dots,Q,
\end{eqnarray}
 \begin{eqnarray}\label{t2_gv1}
v^{(0)}_{\alpha\beta}=
\sum_{\gamma_1=1}^Q  \sum_{\gamma_2=1}^Q 
\Big[\tilde H_{\alpha\gamma_1}'(\rho^{(10)}(w^{(0)})t_1,x_1 I - \rho^{(11)}(w^{(0)}) t_1,\dots,\\\nonumber
x_N I - \rho^{(1N)}(w^{(0)}) t_1  )
\Big]_{\gamma_2 \beta}
 (u^{(0)})_{\gamma_1\gamma_2},
\;\;
\alpha,\beta=1,\dots,Q ,
\end{eqnarray}
where $u^{(0)}=u_1(x)|_{t_2=0}$. 
Eqs.  
(\ref{t2_gu3},\ref{t2_gv1}) must be solved for $u^{(0)}$ and 
$\tilde H_{\alpha\gamma_1}(t_1,z_1,\dots,z_N)$.
 Then eq.(\ref{t2_matr_w22}) gives  
   $w^{(0)}$. 

3.
After that,  the functions  $w$, $u_1$ and $v$  can be constructed as the solutions of  
  the non-differential system (\ref{Sol1_w},\ref{Sol1_u1},\ref{Sol1_v}), $\forall t_1,t_2$.

%%%%%%%%%%%%%%
\section{Auxiliary linear system}
\label{Linear_system}
\label{Linear}
We have seen in the previous sections that the dressing algorithm produces, together with the integrable nonlinear PDEs in 
arbitrary dimensions, also the associated linear overdetermined system of equations for the spectral function $U$, 
whose coefficients are related to the fields of the nonlinear PDEs. 
   
In this section we show that, {\bf unlike the classical $S$-integrable case, the integrability condition for the  
overdetermined system for the spectral function $U$ does not fix completely the integrable nonlinear PDEs}. 
This is not surprising, 
since the derivation of the nonlinear PDEs of this paper requires, together with the linear system for $U$, 
also  the external dressing function $G$ and the constraint (\ref{condition}) for it.

Consider the following linear system, which corresponds to the general form of the nonlinear PDEs 
treated in this paper:
\begin{eqnarray}\label{lin2}
E^{(0)}&:=&A(\lambda,\mu)*U(\mu;x) = U(\lambda;x) w(x),\\\label{lin1}
E^{(m)}&:=&U_{t_m}(\lambda;x) + \sum_{j=1}^N U_{x_j} (\lambda;x) \rho^{(mj)}(x) = U(\lambda;x) F^{(m)}(x)
\end{eqnarray}
for the arbitrary matrix coefficients $w$, $\rho^{(im)}$ and $F^{(m)}$. 
The compatibility of equation (\ref{lin1}) with (\ref{lin2}), for any fixed $m$, yields:
\begin{eqnarray}\label{comp1}
\sum_{j=1}^N U_{x_j} [w, \rho^{(mj)}]  + U
\Big[w_{t_m} + \sum_{j=1}^Nw_{x_j} \rho^{(mj)} - [w,F^{(m)}]\Big] =0.
\end{eqnarray}
Assuming that $U$ and its derivatives are independent matrix functions, equation (\ref{comp1}) implies the 
following relations among the coefficients of the system (\ref{lin1},\ref{lin2}):
  \begin{eqnarray}\label{comp11}
&&   [w, \rho^{(mj)}]=0,\\\label{comp12}
&&   w_{t_m} + \sum_{j=1}^Nw_{x_j} \rho^{(mj)} - [w,F^{(m)}]=0.
  \end{eqnarray}
 
In the scalar case, eq.(\ref{comp11}) is identically satisfied, so that 
 eq.(\ref{comp12}) is some relation among $\rho^{(mj)}$, $w$ and $F^{(m)}$ for any $m$.   
 Thus, the system (\ref{comp11},\ref{comp12}) cannot be considered as a complete system of nonlinear 
equations for some fields.
 
In the matrix case, eq.(\ref{comp11}) implies that the $\rho^{(mj)}$'s are arbitrary functions of $w$, 
representable in a power series of $w$ with scalar coefficients depending arbitrarily on $x$. The 
particular case in which these coefficients are independent of $x$, is
in agreement with the form of the nonlinear PDEs derived in this paper, but no condition is 
imposed on the coefficients $F^{(m)}$. Therefore, also in this case, eqs.(\ref{comp11},\ref{comp12}) cannot be 
 considered as a complete system of nonlinear equations. 
 
The compatibility of two equations from the list (\ref{lin1}), for instance,  $E^{(m)}$ and 
$E^{(n)}$, $n\neq m$,  yields instead:
\begin{eqnarray}
&&
\sum_{i=1}^N \sum_{j=1}^N U_{x_i x_j} [\rho^{(nj)},\rho^{(mi)}] + \\\nonumber
&&
\sum_{j=1}^N U_{x_j}\left[ \sum_{i=1}^N\Big( \rho^{(nj)}_{x_i} \rho^{(mi)} - 
\rho^{(mj)}_{x_i} \rho^{(ni)}\Big) +(\rho^{(nj)}_{t_m} - \rho^{(mj)}_{t_n})-
[F^{(n)},\rho^{(mj)}] + [F^{(m)},\rho^{(nj)}]\right]+\\\nonumber
&&
U\left[
F^{(m)}_{t_n} -F^{(n)}_{t_m} +\sum_{i=1}^N ( F^{(m)}_{x_i} \rho^{(ni)}-F^{(n)}_{x_i} \rho^{(mi)}) +
[F^{(n)},F^{(m)}]\right] = 0.
\end{eqnarray}
Assuming again the independence of $U$ and its derivatives, one obtains the following relations 
among the coefficients:
\beq\label{comp21}
\ba{l}
[\rho^{(nj)},\rho^{(mi)}]=0, \\ 
\rho^{(nj)}_{t_m}+\sum\limits_{i=1}^N\rho^{(nj)}_{x_i}\rho^{(mi)}-[\rho^{(nj)},F^{(m)}]=\;n\leftrightarrow\;m , \\
F^{(n)}_{t_m} +\sum\limits_{i=1}^NF^{(n)}_{x_i}\rho^{(mi)}-F^{(n)}F^{(m)}=\;n\leftrightarrow\;m.
\ea
\eeq
Equation (\ref{comp21}a) follows from the relations (\ref{comp11}), and is satisfied if, for instance,  
$\rho^{(ni)}$ 
are functions of $w$, as it has been previously established. Then equation (\ref{comp21}b) is a consequence of  
 the relation (\ref{comp12}). 
Eq.(\ref{comp21}c) prescribes  relations among the $F^{(m)}$'s, but leaving one of them free.  
Therefore the system (\ref{comp11}-\ref{comp12},\ref{comp21}) cannot be considered as a complete 
system of nonlinear PDEs for the coefficients of the linear system ({\ref{lin2},\ref{lin1}). 

%%%%%%%%%%%%%%%%%%%%
\section{Conclusions}
\label{Conclusions}
 Using a new version of the dressing method, based on a homogeneous integral equation with nontrivial kernel, we 
 constructed a new type of integrable multidimensional nonlinear PDEs. There are no formal restrictions on 
the dimensionality of the PDEs, while these restrictions are very severe in the case of the classical 
$S$-integrable systems, which are known to be   the first examples of nonlinear PDEs treatable by 
dressing technics. 
 Several modifications and extensions of the dressing algorithm presented here will be considered in 
subsequent papers. 
 
 \vskip 10pt \noindent {\bf Acknowledgments}. 
AIZ was supported by the INTAS Young Scientists Fellowship
Nr. 04-83-2983, by the RFBR grants 04-01-00508, 06-01-90840,
06-01-92053 and by
the grant Ns 7550.2006.2. 
 
%%%%%%%%%%%%%%%%%%%%%%%%%%%%%


\begin{thebibliography}{100}

\bibitem{GGKM}
C.S.Gardner, J.M.Green, M.D.Kruskal, R.M.Miura, Phys.Rev.Lett,
{\bf 19},  (1967) 1095
 
\bibitem{KdV} 
D. J. Korteweg and G. de Vries, Philos. Mag. Ser. 5, {\bf 39},
(1895) 422 

\bibitem{ZMNP}
V.E.Zakharov, S.V.Manakov, S.P.Novikov and L.P.Pitaevsky, 
{\it Theory of Solitons. The Inverse Problem Method}, 
Plenum Press (1984)

\bibitem{AS}
M.J.Ablowitz and P.C.Clarkson, {\it Solitons, Nonlinear Evolution Equations and Inverse Scattering}, 
Cambridge University Press, Cambridge, 1991

\bibitem{Calogero}
F.Calogero in {\it What is integrability} by V.E.Zakharov,
Springer,  (1990) 1

\bibitem{ZSh1}
V.E.Zakharov and A.B.Shabat, Funct.Anal.Appl., {\bf 8}, (1974) 43 
\bibitem{ZSh2}
V.E.Zakharov and A.B.Shabat, Funct.Anal.Appl., {\bf 13},  (1979) 13 

\bibitem{ZM}
V.E.Zakharov and S.V.Manakov, Funct.Anal.Appl., {\bf 19}, (1985) 11 

\bibitem{BM}
L.V.Bogdanov and S.V.Manakov, J.Phys.A:Math.Gen., {\bf 21},
(1988) L537 

\bibitem{Konop}
B. Konopelchenko, {\it Solitons in Multidimensions}, World Scientific, Singapore (1993)

\bibitem{Zakharov1}
V. E. Zakharov,
Lecture Notes in Physics, {\bf 153}, 
Springer-Verlag , Berlin, (1982) 190 

\bibitem{Zakharov2}
V. E. Zakharov,
in Proceedings of the International Congress of Mathematicians, 
PWN Warsaw, (1983) 1225 



\bibitem{Zakharov}
V. E. Zakharov,  in Inverse Methods in Action, 
edited by P. C. Sabatier, Springer-Verlag, Berlin, (1990) 602 



\bibitem{Z}
 A.Zenchuk,
 J.Physics A:Math.Gen. {\bf 37}, nn 25, (2004) 6557 


\bibitem{BK}
L.V.Bogdanov and B.G.Konopelchenko, Phys.Lett.A, {\bf 345},
(2005) 137

\bibitem{MS}
S.V.Manakov and P.M.Santini,  Phys. Lett. A {\bf 359}, (2006) 613

\bibitem{ZS} A.I.Zenchuk, P.M.Santini, J. Phys. A: Math. Gen. {\bf 39} (2006) 
5825

\bibitem{Z3} A.I.Zenchuk, arXiv: nlin.SI/0612048 

\bibitem{Z2}
 A.I.Zenchuk, math.AP/0603294 


\bibitem{Whitham} J. B. Whitham, {\it Linear and Nonlinear Waves}, Wiley, NY, 1974

\bibitem{SZ} P.M.Santini and A.I.Zenchuk, arXiv:nlin.SI/0612036 


\bibitem{ts1}
S.P.Tsarev, 
{\it Soviet Math. Dokl.}, {\bf 31},  n. 3, (1985)
488

\bibitem{dn}
B.A. Dubrovin and S.P. Novikov,
{\it Russian Math. Survey}, {\bf 44} n.6 (1989)
35

\bibitem{ts2}
S.P.Tsarev, Math. USSR Izvestiya, {\bf 37} (1991) 397


\bibitem{SAF} P.M.Santini, M.J.Ablowitz and A.S.Fokas, J.Math.Phys. 25, 2614  (1984).

\bibitem{BLR} M. Bruschi, D. Levi and O. Ragnisco, Il Nuovo Cimento B {\bf 74},  (1983) 33

\end{thebibliography}
\end{document}